\documentclass[prd,superscriptaddress,amsfonts,amssymb,amsmath,showpacs,twocolumn]{revtex4-1}
\usepackage{textcomp}
\usepackage{eurosym}
\usepackage{amsfonts}
\usepackage{array}
\usepackage{amsthm}
\usepackage{bm}
\usepackage[colorinlistoftodos]{todonotes}
\usepackage{mathpazo}
\usepackage{supertabular}
\usepackage{subfig}
\usepackage{amssymb}
\usepackage{eurosym}
\usepackage{amsmath}
\usepackage{epsfig}
\usepackage{graphics}
\usepackage{color}
\usepackage{graphicx}
\usepackage[font={footnotesize,it}]{caption}
\usepackage[colorlinks=true,
            linkcolor=blue,
            urlcolor=blue,
            citecolor=green]{hyperref}
\begin{document}
\title{Traversable Wormholes Supported by GUP Corrected Casimir Energy}

\author{Kimet Jusufi}
\email{kimet.jusufi@unite.edu.mk}
\affiliation{Physics Department, State University of Tetovo, Ilinden Street nn, 1200,
Tetovo, Macedonia}
\affiliation{Institute of Physics, Faculty of Natural Sciences and Mathematics, Ss. Cyril and Methodius University, Arhimedova 3, 1000 Skopje, Macedonia}

\author{Phongpichit Channuie} 
\email{channuie@gmail.com}
\affiliation{School of Science, Walailak University, Thasala, \\Nakhon Si Thammarat, 80160, Thailand}
\affiliation{College of Graduate Studies, Walailak University, Thasala, Nakhon Si Thammarat, 80160, Thailand}
\affiliation{Research Group in Applied, Computational and Theoretical Science (ACTS), Walailak University, Thasala, Nakhon Si Thammarat, 80160, Thailand}
\affiliation{Thailand Center of Excellence in Physics, Ministry of Education, Bangkok 10400, Thailand}

\author{Mubasher Jamil}
\email{mjamil@zjut.edu.cn (corresponding author)}
\affiliation{Institute for Theoretical Physics and Cosmology, Zhejiang University of Technology, Hangzhou, 310023 China}
\affiliation{Department of Mathematics, School of Natural Sciences (SNS), National University of Sciences and Technology (NUST), H-12, Islamabad, 44000 Pakistan}

\begin{abstract}
In this paper, we investigate the effect of the Generalized Uncertainty Principle (GUP) in the Casimir wormhole spacetime recently proposed by Garattini [Eur. Phys. J. C (2019) 79: 951]. In particular, we consider three types of the GUP relations, firstly the Kempf, Mangano and Mann (KMM) model, secondly the Detournay, Gabriel and Spindel (DGS) model, and finally the so called type II model for GUP principle. To this end, we consider three specific models of the redshift function along with two different EoS of state given by  $\mathcal{P}_r(r)=\omega_r(r) \rho(r)$ along with $\mathcal{P}_t(r)=\omega_t (r)\mathcal{P}_r(r)$ and obtain a class of asymptotically flat wormhole solutions supported by Casimir energy under the effect of GUP. Furthermore we check the null, weak, and strong condition at the wormhole throat with a radius $r_0$, and show that in general the classical energy condition are violated by some arbitrary quantity at the wormhole throat. Importantly, we examine the wormhole geometry with semi-classical corrections via embedding diagrams. We also consider the ADM mass of the wormhole, the volume integral quantifier to calculate the amount of the exotic matter near the wormhole throat, and the deflection angle of light.
\end{abstract}
\pacs{}
\keywords{}
\maketitle

\section{Introduction}

The search for a theory of exotic objects through Einstein’s general theory of relativity has been receiving a lot of interest in the literature. A black hole, e.g. the Schwarzschild black
hole, possesses one of the possible solutions to Einstein’s field equations, see Ref.\cite{Schwz}. The recent detection of gravitational waves (GWs) \cite{Abbott:2016blz} demonstrated that stellar-mass black holes really exist in Nature. Interestingly, the author of Ref.\cite{Flamm} realized in 1916 that another solution was viable which is presently known as a “white hole”. 

In 1935, Einstein and Rosen used the theory of general relativity to propose the existence of "bridges" through space-time \cite{ERb}. These bridges connect two different points in space-time enable to create a shortcut called Einstein-Rosen bridges, or wormholes. However, the existence of wormholes needs to be experimentally observed. Moreover, Morris and Throne \cite{Morris} demonstrated that wormholes are solutions of Einstein field equations. Hypothetically, they connect two space-time regions of the universe by a throat. The first type of wormhole solution was the Schwarzschild wormhole \cite{Vladimir} which would be present in the Schwarzschild metric describing an eternal black hole. However, it was found that it would collapse too quickly. In principle, it is possible to stabilize the wormholes if there exists an exotic matter with negative energy density.

In order to maintain the structure of the wormhole, we need the exotic matter which satisfies the flare-out condition and violates
weak energy condition \cite{Morris:1988cz,Morris:1988tu}. Classically, there are no traversable wormholes. However, it has been recently shown that quantum matter fields can provide enough negative
energy to allow some wormholes to become traversable. As a result, to construct such a traversable wormhole, one requires an exotic matter with a negative energy density and a large negative pressure, which should have a
higher value than the energy density.

In the literature, many authors have intensively studied various aspects of traversable wormhole (TW) geometries
within different modified gravitational theories
\cite{Harko:2013aya,Lobo:2009ip,Bohmer:2011si,Zangeneh:2015jda,Clement:1983fe,Bronnikov:2010tt,Visser:2003yf,Mehdizadeh:2017tcf,Jusufi:2017drg,Jusufi:2016leh,Jusufi:2018waj,Jusufi:2019knb,Dai:2019mse,Tsukamoto:2014swa,Tsukamoto:2016zdu,Tripathi:2019trz,Teo:1998dp,Ovgun:2018xys,Shaikh:2018yku,Shaikh:2016dpl,Ovgun:2018prw,MontelongoGarcia:2010xd,Rahaman:2016jds,Rahaman:2014dpa,Jamil:2013tva,Rahaman:2012pg,Sahoo:2017ual}. Recently the shadows of wormholes and Kerr-like wormholes was  investigated in Refs.\cite{Shaikh:2018kfv,Amir:2018pcu,Gyulchev:2018fmd,Jusufi:2018gnz,Amir:2018szm,Tsukamoto:2012xs}.
These include $f(R)$ and $f(T)$ theories, see e.g., \cite{Bahamonde:2016jqq,Bahamonde:2016ixz,Jamil:2012ti,Jamil:2008wu}. Among them, as the possibility of phantom energy, this presents us with a natural
scenario for the existence of traversable wormholes \cite{Lobo:2005us}. In addition, the wormhole construction
in $f(R)$ gravity is studied in Refs.\cite{Bahamonde:2016ixz,Rahaman:2013qza}.  Interestingly, the Casimir effect also provides a possibility to produce negative energy density and can be used to stabilze tranversable wormholes.

The main aim of this paper is to investigate the effect of the Generalized Uncertainty Principle (GUP) in the Casimir Wormhole spacetime recently proposed by Garattini \cite{Garattini:2019ivd}. In particular, we consider three types of the GUP relations: (1) the Kempf, Mangano and Mann (KMM) model, (2) the Detournay, Gabriel and Spindel (DGS) model, and (3) the so called type II model for GUP principle. We study a class of asymptotically flat wormhole solutions supported by Casimir energy under the effect of GUP.

This paper is organized as follows: In Sec.\ref{Sec2}, we take a short recap of the Casimir effect under the Generalized
Uncertainty Principle and consider three models with the generic functions $f\left(\hat{p}^{2}\right)$ and $g\left(\hat{p}^{2}\right)$. In Sec.\ref{gupWorm}, we construct the GUP Casimir wormholes by particularly focusing on three types of the GUP relations. We then examine the energy conditions of our proposed models in Sec.\ref{Ener} and quantifies the amount of  exotic  matter  required  for  wormhole  maintenance in Sec.\ref{amont}. Furthermore, we study the  gravitational lensing  effect  in  the  spacetime  of  the  GUP  Casimir wormholes in Sec.\ref{chvi}. We finally conclude our findings in the last section. In this present work, we use the geometrical units such that $G=c=1$.

\section{The Casimir Effect under the Generalized
Uncertainty Principle}
\label{Sec2}
The Casimir effect manifests itself as the interaction of a pair of neutral, parallel conducting planes caused by the disturbance of the vacuum 
of the electromagnetic field. The Casimir effect can be described in terms of the zero-point energy of a quantized field in the intervening space between the objects. It is a macroscopic quantum effect which causes the plates to attract each other. In his famous paper \cite{Casimir}, Casimir derived the finite energy between plates and found that the energy per unit surface is given by
\begin{eqnarray}
\mathcal{E}=-\frac{\pi^{2}}{720}\frac{\hbar }{a^{3}},\label{eq:2}
 \end{eqnarray}
where $a$ is a distance between plates along the $z$-axis, the direction perpendicular to the plates. Consequently, we can determine the finite force per unit area acting between the plates to yield $\mathcal{F}=-\frac{\pi^{2}}{240}\frac{\hbar }{a^{4}}$. Notice that the minus sign corresponds to an attractive force. The resolution of small distances in the spacetime is limited by the existence of a minimal length in the theory. Note that the prediction of a minimal measurable length in order of Planck length in various theories
of quantum gravity restricts the maximum energy that any particle can attain to the Planck
energy. This implied the modification of linear momentum and also quantum commutation
relations and results the modified dispersion relation, e.g., gravity's rainbow \cite{Magueijo:2002xx}, see some particular cosmological \cite{Chatrabhuti:2015mws,Channuie:2019kus,Hendi:2016tiy} and astrophysical implications \cite{Hendi:2016hbe,Feng:2017gms,Hendi:2018sbe,Panahiyan:2018fpb,Dehghani:2018qvn}. Moreover, this scale naturally arises in theories of quantum gravity in the form of an effective minimal uncertainly in positions $\Delta x_{0}>0$. 

For instance, in string theory, it is impossible to improve the spatial resolution below the characteristic length of the strings. As a results, a correction to the position-momentum uncertainty relation related to this characteristic length can be obtained. In one dimension, this minimal length can be implemented adding corrections to the uncertainty relation to obtain
\begin{eqnarray}
\label{MUC}
\Delta x \Delta p \geq \frac{\hbar}{2}\left[1+\beta\left(\Delta p\right)^{2}+\gamma\right],\qquad\beta,\gamma>0, \label{cr}
\end{eqnarray}
where a finite minimal uncertainty $\Delta x_{0}=\hbar\sqrt{\beta}$ in terms of the minimum length parameter $\beta$ appears. As a result, the modification of the uncertainty relation Eq. \eqref{cr} implies a small correction term to the usual Heisenberg commutator relation of the form:
\begin{eqnarray}
\label{MCR}
\left[\hat{x},\hat{p}\right]=i\hbar\left(1+\beta\hat{p}^{2}+\ldots\right). \label{H}
\end{eqnarray}
It is worth noting in these theories that the eigenstates of the position operator are no longer physical states whose matrix elements would have the usual direct physical interpretation about positions. Therefore, one introduces the "quasi-position representation", which consists in projecting the states onto the set of maximally localized states. Interestingly, the usual commutation relation given in Eq.(\ref{H}) can be basically generalized. In $n$ spatial dimensions, the generalized commutation relations leading to the GUP that provides a minimal uncertainty are assumed of the form \cite{Frassino:2011aa}:
\begin{eqnarray}
\left[\hat{x}_{i},\,\hat{p}_{j}\right] & = & i\hbar\left[f\left(\hat{p}^{2}\right)\delta_{ij}+g\left(\hat{p}^{2}\right)\hat{p}_{i}\hat{p}_{j}\right]\,, \label{RC}
\end{eqnarray}
where $i,j=1, ... n$ and the generic functions $f\left(\hat{p}^{2}\right)$ and $g\left(\hat{p}^{2}\right)$ are not necessarily arbitrary. Note that the relations between them can be quantified by imposing translational and rotational invariance on the generalized commutation relations. As mentioned in Ref.\cite{Frassino:2011aa}, the specific form of these states depends on the number of dimensions and on the specific model considered. For example, when $n>1$ the generalized uncertainty relations are not unique and different models may be obtained by choosing different functions $f\left(\hat{p}^{2}\right)$ and/or  $g\left(\hat{p}^{2}\right)$ which will yield  different maximally localized states.

\subsection{Model I (KMM)}
 The specific form of these states depends on the number of dimensions and on the specific model considered. In 
literature there are at least two different approaches to construct maximally localized states: the procedure proposed by Kempf, Mangano and Mann (KMM). This model correspond to  the choice of the generic functions $f\left(\hat{p}^{2}\right)$ and $g\left(\hat{p}^{2}\right)$ given in  Ref. \cite{Frassino:2011aa}:
\begin{eqnarray}f\left(\hat{p}^{2}\right)=\frac{\beta\hat{p}^{2}}{\sqrt{1+2\beta\hat{p}^{2}}-1},\qquad g\left(\hat{p}^{2}\right)=\beta\,.\label{eq:4}\end{eqnarray}
 From now on we will remove the hat over the operator.
Following the KMM construction, one obtains then the final result with the first order correction term in the minimal uncertainty parameter $\beta$ introduced in the modified  commutation relations of Eq.~\eqref{MCR} : 
\begin{eqnarray}
 \mathcal{E} & = & -\frac{\pi^{2}}{720}\frac{\hbar }{a^{3}}\left[1+\pi^{2}\left(\frac{28+3\sqrt{10}}{14}\right)\left(\frac{\hbar\sqrt{\beta}}{a}\right)^{2}\right]. \label{eq:RisultatoK}
 \end{eqnarray}
The force per unit area  relation in this model is given by
\begin{eqnarray}
 \mathcal{F} & = & -\frac{\pi^{2}}{240}\frac{\hbar }{a^{4}}\left[1+\pi^{2}\left(\frac{10}{3}+\frac{5\sqrt{10}}{14}\right)\left(\frac{\hbar\sqrt{\beta}}{a}\right)^{2}\right]. \label{e7}
 \end{eqnarray}

\subsection{Model I (DGS)}  
In this model the Casimir energy per unit surface is given by \cite{Frassino:2011aa}
\begin{eqnarray} 
  \mathcal{E}& = & -\frac{\pi^{2}}{720}\frac{\hbar }{a^{3}}\left[1+\pi^{2}\frac{4\left(3+\pi^{2}\right)}{21}\left(\frac{\hbar\sqrt{\beta}}{a}\right)^{2}\right]\,.
 \label{eq:RisultatoK}
\end{eqnarray}
On the other hand, the finite force per unit area acting between the plates 
\begin{eqnarray}
 \mathcal{F}& = & -\frac{\pi^{2}}{240}\frac{\hbar }{a^{4}}\left[1+\pi^{2}\left(\frac{20}{21}+\frac{20\pi^{2}}{63}\right)\left(\frac{\hbar\sqrt{\beta}}{a}\right)^{2}\right]. \label{e9}
 \end{eqnarray}

\subsection{Model II } 
The model proposed is completely different from that given by Eq. \eqref{eq:4}. This model has 
the functions $f$ and $g$ as follows:\cite{Frassino:2011aa}
\begin{eqnarray}
f\left(p^{2}\right)=1+\beta p^{2},\qquad g\left(p^{2}\right)=0.
\end{eqnarray}
One obtains then the  final result \cite{Frassino:2011aa}
\begin{eqnarray}
\mathcal{E} 
  =  -\frac{\pi^{2}}{720}\frac{\hbar }{a^{3}}\left[1+\pi^{2}\frac{2}{3}\left(\frac{\hbar\sqrt{\beta}}{a}\right)^{2}\right].\label{eq:Risultato}
\end{eqnarray}
The first term in Eq. \eqref{eq:Risultato} is the usual Casimir energy reported in Eq. \eqref{eq:2} and is obtained without the cut-off function. 
The second term is the correction given by the presence in the theory of a minimal length. We note that it is attractive.
The force per unit area  in this model is given by
\begin{eqnarray}
 \mathcal{F} & = & -\frac{\pi^{2}}{240}\frac{\hbar }{a^{4}}\left[1+\pi^{2}\frac{10}{9}\left(\frac{\hbar\sqrt{\beta}}{a}\right)^{2}\right]. \label{e12}
 \end{eqnarray}

\subsection{GUP corrected energy density}
Let us know elaborate in more details about the GUP corrected energy densities by writing first the renormalized energies for three GUP cases
\begin{equation}
E=-\frac{\pi^{2}S}{720}\frac{\hbar }{a^{3}}\left[1+C_i\left(\frac{\hbar\sqrt{\beta}}{a}\right)^{2}\right]
\end{equation}
where $S$ is the surface area of the plates and
$a$ is the separation between them. Note that we have introduced the constant $C_i$ where $i=1,2,3$. In particular we have the following three cases:
\begin{eqnarray}
    C_1&=&\pi^{2}\left(\frac{28+3\sqrt{10}}{14}\right),\\
    C_2&=&4\pi^{2}\left(\frac{3+\pi^2}{21}\right),\label{e19}\\
    C_3&=&\frac{2 \pi^2}{3}.\label{e20}
\end{eqnarray}

Then the force can be obtained with the
computation of 
\begin{equation}
F=-\frac{dE}{da}=-\frac{3 \pi^{2}S}{720}\frac{\hbar }{a^{4}}\left[1+\frac{5}{3}C_i\left(\frac{\hbar\sqrt{\beta}}{a}\right)^{2}\right]
\end{equation}

Thus, using 
\begin{equation}
P=\frac{F}{S}=-\frac{3 \pi^{2}}{720}\frac{\hbar }{a^{4}}\left[1+\frac{5}{3}C_i\left(\frac{\hbar\sqrt{\beta}}{a}\right)^{2}\right]=\omega \rho.
\end{equation}

At this point we note that in the case of Casimir energy there is a natural EoS establishing fundamental relationship by choosing $\omega=3$. From the last equation we obtain the GUP corrected energy density in a compact form as 
\begin{equation}
\rho=-\frac{ \pi^{2}}{720}\frac{\hbar }{a^{4}}\left[1+\frac{5}{3}C_i\left(\frac{\hbar\sqrt{\beta}}{a}\right)^{2}\right].
\end{equation}
Setting $\beta=0$, we obtain the usual Casimir result. In this way we can introduce a new constant $D_i=5C_i/3$, however
the GUP extension seems to be not uniquely defined, therefore different extensions lead to different $D_i$. This, on the other hand, suggests a possible  extension of energy density. For example, one can postulate the following extension
\begin{equation}
\rho=-\frac{ \pi^{2}}{720}\frac{\hbar }{a^{4}}\left[1+A_i\left(\frac{\hbar\sqrt{\beta}}{a}\right)^{2}+B_i\left(\frac{\hbar\sqrt{\beta}}{a}\right)^{4}+... \right],
\end{equation}
where $A_i$ and $B_i$ are some constants. In the present work we shall use the expression (19)  for the energy density and leave Eq. (20) for future work.

\section{GUP Casimir Wormholes}
\label{gupWorm}
We consider a static and spherically symmetric Morris-Thorne traversable wormhole in the Schwarzschild coordinates given by \cite{Morris} 
\begin{equation}
\mathrm{d}s^{2}=-e^{2\Phi (r)}\mathrm{d}t^{2}+\frac{\mathrm{d}r^{2}}{1-\frac{b(r)}{r}}+r^{2}\left(
\mathrm{d}\theta ^{2}+\sin ^{2}\theta \mathrm{d}\phi ^{2}\right),  \label{5}
\end{equation}
in which $\Phi (r)$ and $b(r)$ are the redshift and shape
functions, respectively. In the wormhole geometry, the redshift
function $\Phi (r)$ should be finite in order to avoid the
formation of an event horizon. Moreover, the shape function $b(r)$ determines the wormhole geometry, with the following condition $b(r_{0})=r_{0}$, in which $r_{0}$ is the radius of the wormhole throat. Consequently, the shape function must satisfy the flaring-out condition \cite{Morris}: 
\begin{equation}
\frac{b(r)-rb^{\prime }(r)}{b^{2}(r)}>0, 
\end{equation}%
in which $b^{\prime }(r)=\frac{db}{dr}<1$ must hold at the throat of the
wormhole. With the help of the line element \eqref{5}, we obtain the following set of equations resulting from 
the energy-momentum components to yield
\begin{eqnarray}
\rho (r) &=&\frac{1}{8\pi r^{2}} b^{\prime }(r), \\
\mathcal{P}_{r}(r) &=&\frac{1}{8\pi }\left[ 2\left( 1-\frac{b(r)}{r}\right) 
\frac{\Phi ^{\prime }}{r}-\frac{b(r)}{r^{3}}\right]
,  \\
\mathcal{P}_t(r) &=&\frac{1}{8\pi }\left( 1-\frac{b(r)}{r}\right) \Big[\Phi
^{\prime \prime }+(\Phi ^{\prime })^{2}-\frac{b^{\prime }r-b}{2r(r-b)}\Phi
^{\prime }  \notag \\
&-&\frac{b^{\prime }r-b}{2r^{2}(r-b)}+\frac{\Phi ^{\prime }}{r%
}\Big].  \label{18}
\end{eqnarray}%
where $\mathcal{P}_t=\mathcal{P}_{\theta }=\mathcal{P}_{\phi }$. 

Having used the energy density, we can find shape function $b(r)$ and then we can use the EoS with a specific value for $\omega$ to determine the redshift function. However, in general, it is known that most of the solutions are unbounded if $r$ is very large. Hence such corresponding solutions may not be physical. In the present paper, we are interested in deriving the equation of state (connecting pressures with density) for a given wormhole geometry. In other words, we fix the geometry parameters using different redshift functions of a wormhole and then ask what the EoS parameter in the corresponding case is. Moreover, we also need to check the behavior of energy conditions near the throat. 
In order to simplify the notation from now on, we shall set the Planck constant to one, i.e., $\hbar=1$.

\subsection{Model $\Phi=constant$}
To simplify our calculations, we are going to introduce $D_i$ and the replacement $a \to r$ in the expression for the energy density. In that case, using Eq. (19) the energy density relations can be rewritten 
\begin{eqnarray}
 \rho& = & -\frac{\pi^{2}}{720 r^4}\left[1+D_{i}\left(\frac{\sqrt{\beta}}{r}\right)^{2}\right]. \label{e18}
 \end{eqnarray}
where $i=1,2,3$. In particular we have the following three cases:
\begin{eqnarray}
    D_1&=&5\,\pi^{2}\left(\frac{28+3\sqrt{10}}{42}\right),\\
    D_2&=&20\,\pi^{2}\,\left(\frac{3+\pi^2}{63}\right),\label{e19}\\
    D_3&=&\frac{10 \pi^2}{9}.\label{e20}
\end{eqnarray}
The simplest case is a model with $\Phi=constant$, namely a spacetime with no tidal forces, namely $\Phi'(r)=0$. In other words, this is asymptotically flat wormhole spacetime. 
We find
\begin{equation}
b(r)=C_1+\frac{\pi^3 }{90 r}+\frac{\pi^3  D_{i} \beta}{270 r^3}.
\end{equation}

\begin{figure}
\includegraphics[width=8.2cm]{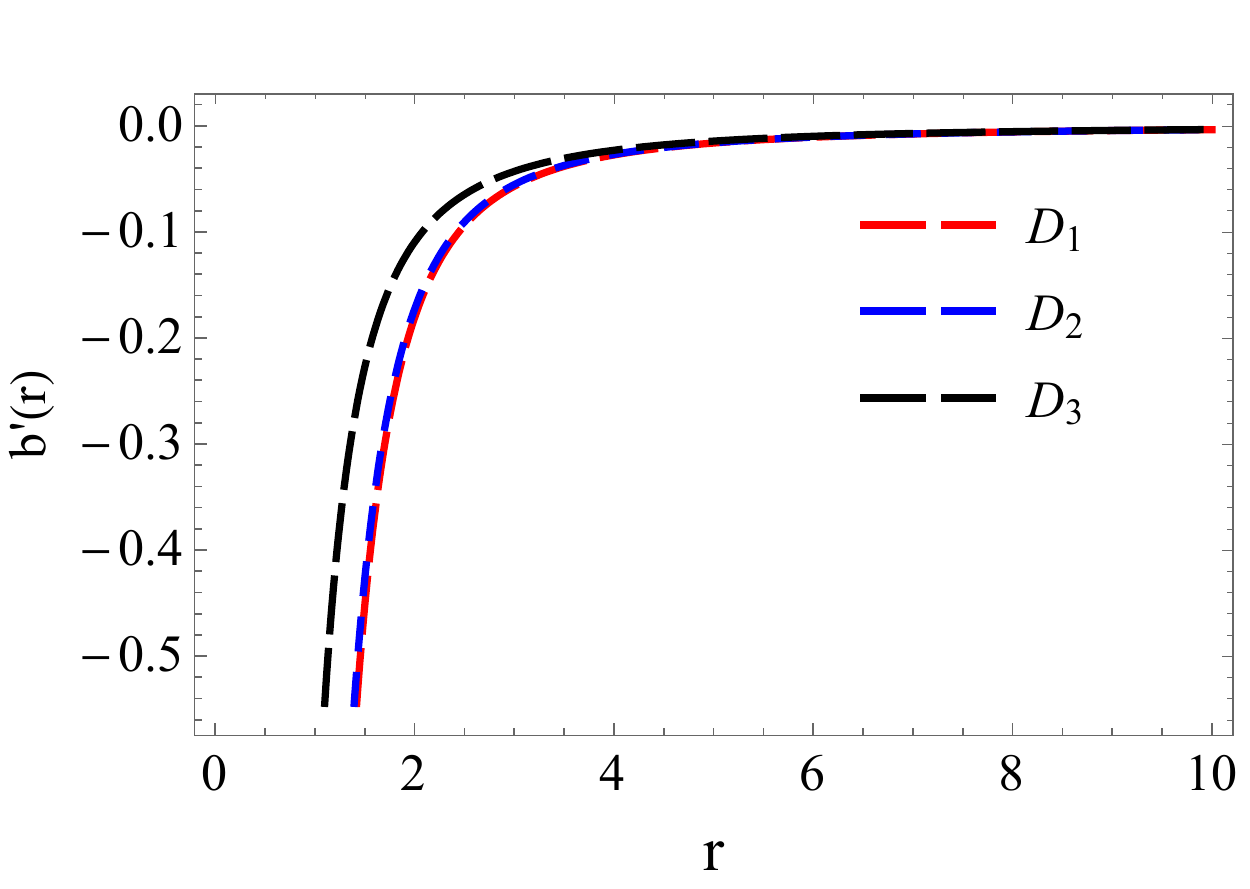}
\caption{ We check the flare out condition. Variation of $b'(r)$ against $r$. We have used $r_0=1$, $\hbar=1$ and $\beta=0.1$.  }\label{fig1}
\end{figure}

\begin{figure}
\includegraphics[width=8.2cm]{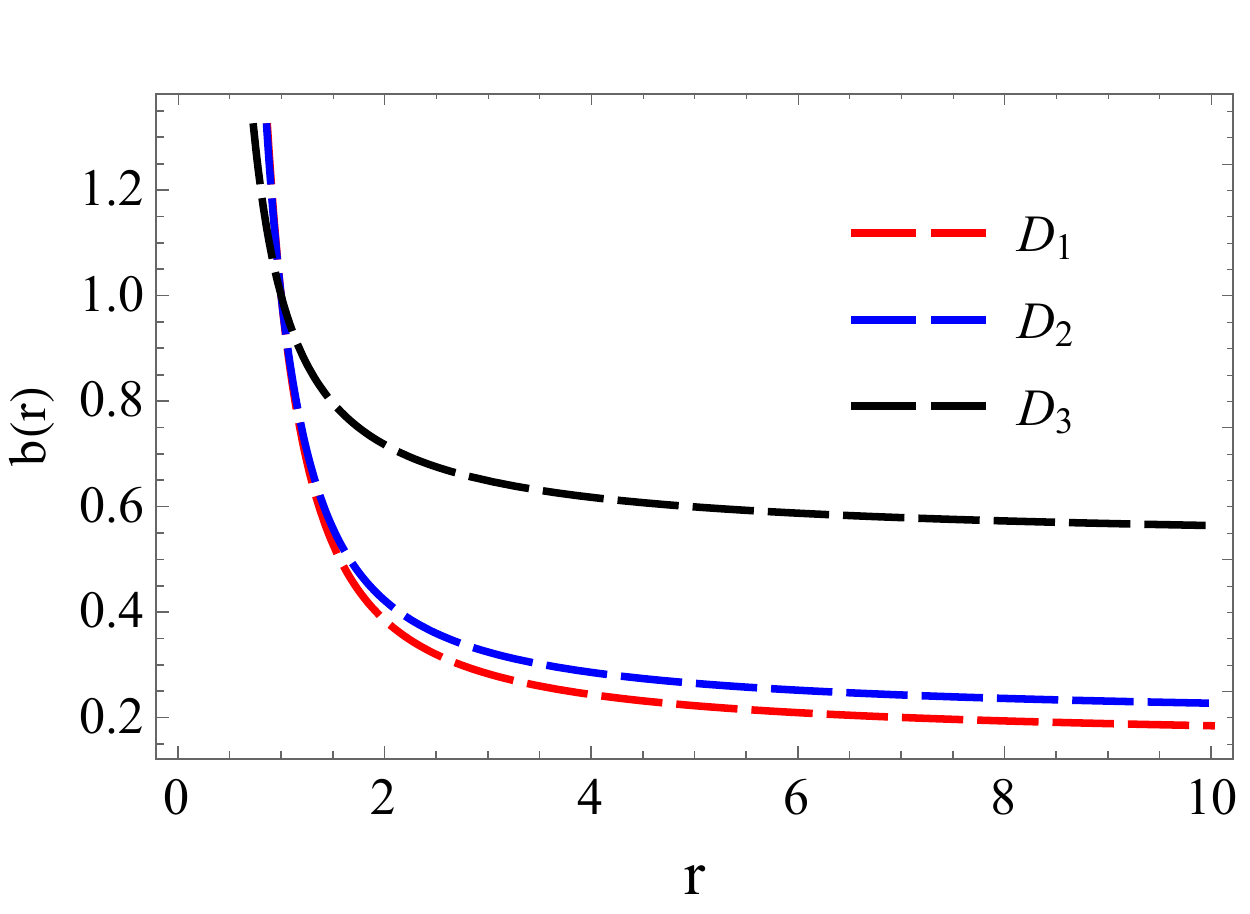}
\caption{The shape function of the GUP wormhole against $r$. We use $\hbar=1$ and $\beta=0.1$.  }
\end{figure}
Finally we use $b(r_0)=b_0=r_0$, to calculate the the constant $C$. Thus by solving the last differential equation we find the shape function to be
\begin{equation}
b(r)=r_0+\frac{\pi^3 }{90}\left(\frac{1}{r}-\frac{1}{r_0}\right)+\frac{\pi^3 D_{i} \beta}{270} \left(\frac{1}{r^3}-\frac{1}{r_0^3}\right).\label{s22}
\end{equation}
Introducing the scaling of coordinate $\exp(2 \Phi) dt^2 \to dt^2$ (since $\exp(2 \Phi)=const$), the wormhole metric reads 
\begin{eqnarray}\notag
ds^2&=&-dt^2+\frac{dr^2}{1-\frac{r_0}{r}-\frac{\pi^3}{90 r}\left(\frac{1}{r}-\frac{1}{r_0}\right)-\frac{\pi^3 D_{i} \beta}{270 r} \left(\frac{1}{r^3}-\frac{1}{r_0^3}\right)}\\
&+&r^2 (d\theta^2+\sin^2\theta \,d\phi^2),
\end{eqnarray}
Clearly in the limit $r \to \infty $, we obtain
\begin{equation}
\lim_{ r \to \infty} \frac{b(r)}{r} \to 0.
\end{equation}

The asymptotically flat metric can be seen also from the Fig.(\ref{fig1}). Using the EoS $\mathcal{P}_r(r)=\omega(r) \rho(r)$, one can easily see that when $\Phi(r)=0$ (tideless wormholes) we obtain 
\begin{equation}\label{24}
8 \omega(r) \rho(r) \pi r^3+b(r)=0.
\end{equation}
Solving this equation  for the EoS parameter we obtain
\begin{equation}
    \omega=-\frac{\beta D_{i} \pi^3  (r^3-r_0^3)+3r_0^2r^2((\pi^3-90r_0^2)r-\pi^3 r_0) }{3 (D_{i} \beta +r^2) \pi^3 r_0^3 }.\label{e27}
\end{equation}

\begin{figure}
\includegraphics[width=8.2cm]{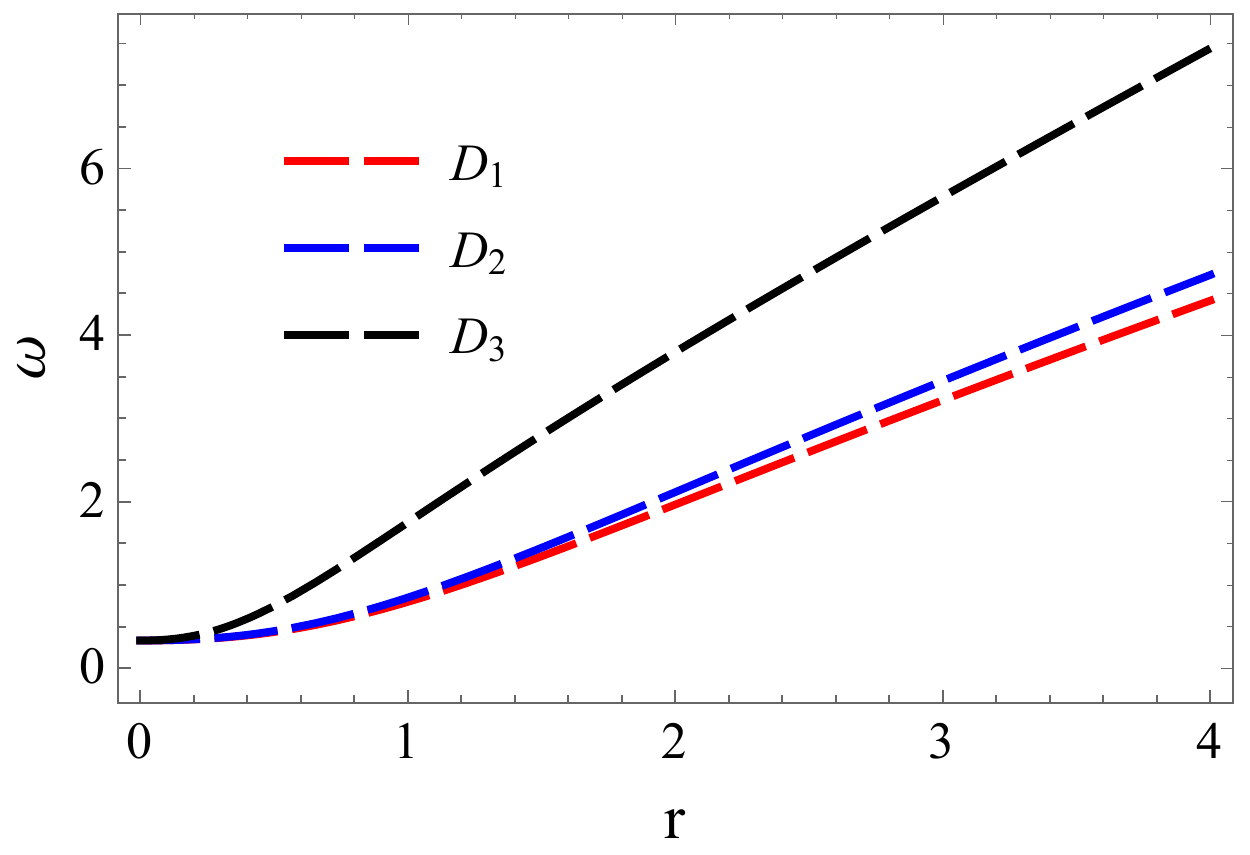}
\caption{The EoS parameter $\omega$ for the GUP wormhole with $\Phi=0$ as a function of $r$. We use $r_0=1$, $\hbar=1$ and $\beta=0.1$. }
\end{figure}

\subsection{Model with $\Phi(r)=\frac{r_0}{r}$}
\subsubsection{EoS: $\mathcal{P}_r(r)=\omega_r(r) \rho(r)$}
We shall begin our analysis by considering the following EoS $\mathcal{P}_r(r)=\omega_r(r) \rho(r)$. From the Einstein's field equations (\ref{18}), we find
\begin{equation}\label{24}
\Phi'(r)=-\frac{8 \omega_r(r) \rho(r) \pi r^3+b(r)}{2 r(-r+b(r))}.
\end{equation}
Now considering the model function
\begin{equation}
    \Phi(r)=\frac{r_0}{r},\label{e26}
\end{equation}
we obtain the following equation
\begin{eqnarray}\notag
\frac{(r-2 r_0)b(r)+8 \omega_1(r) \rho(r) r^4 \pi+2 r_0 r}{8 \pi r^4}=0.
\end{eqnarray}
Finally using the shape function (\ref{e19}) for the EoS parameter we obtain
\begin{equation}
    \omega_r(r)=-\frac{\beta \left[D_{i} \pi^3 (r^4-2r^3 r_0-r r_0^3+2 r_0^4) \right]+\mathcal{F}}{3 (D_{i} \beta +r^2)r \pi^3 r_0^3 },\label{e27}
\end{equation}
where
\begin{eqnarray}
    \mathcal{F}&=&3 \pi^3  r^4 r_0^2-9 \pi^3 r^3 r_0^3+6 \pi^3  r^2 r_0^4-810 r^4 r_0^4\nonumber\\&&+540 r^3 r_0^5.\label{e31}
\end{eqnarray}

\subsubsection{EoS: $\mathcal{P}_t(r)=\omega_t (r) \mathcal{P}_r(r)$}
Let us now consider the scenario in which the EoS is of the form $\mathcal{P}_t(r)=\omega_t(r) \mathcal{P}_r(r)$, where $\omega_t(r)$ is as an arbitrary function of $r$ . In this case, combining the second and the third equation in (\ref{18}) we find the following equation: 
\begin{eqnarray}\notag
   &2& r(r+1)(r-b(r))\Phi''(r)+2r^2 (r-b(r))(\Phi'(r))^2\\\notag
   &-&r \Phi'(r)\left[ (-4 \omega_2(r)+r-1)b(r)+4 \omega_2(r) r \right]\\
   &+&b(r)(2 \omega_2(r) -r +1)=0.
\end{eqnarray}
Using the shape function (\ref{s22}) along with Eq. (\ref{e26}) from the last equation we obtain 
\begin{equation}
   \omega_t(r)= \frac{(r-r_0)\Big[\beta D_{i} \pi^3(r^2+r r_0+r_0^2)\mathcal{H}+3 r_0^2 r^2\mathcal{G}\Big]}{2 r \left[\beta D_{i} \pi^3  (r^4-2 r^3 r_0-r r_0^3+2 r_0^4)+\mathcal{F}\right]},\label{e30}
\end{equation}
where
\begin{eqnarray}
  \mathcal{H} &=&2 r_0^2+r_0(4+5 r-r^2)+r^2(r-1),\label{e31}\\
    \mathcal{G}&=&180 r r_0^3+r_0^2(2  \pi^3-90 r^3+450 r^2+360 r)- \pi^3
    \nonumber\\&&\times r_0 (r^2-5 r-4)+r^2 \pi^3(r-1),\label{e32}\\
    \mathcal{F}&=& 3 \pi^3 r^4 r_0^2-9\pi^3  r^3 r_0^3+6\pi^3  r^2 r_0^4-810 r_0^4 r^4\nonumber\\&&+540 r^3 r_0^5.
\end{eqnarray}

\begin{figure}
\includegraphics[width=8.2cm]{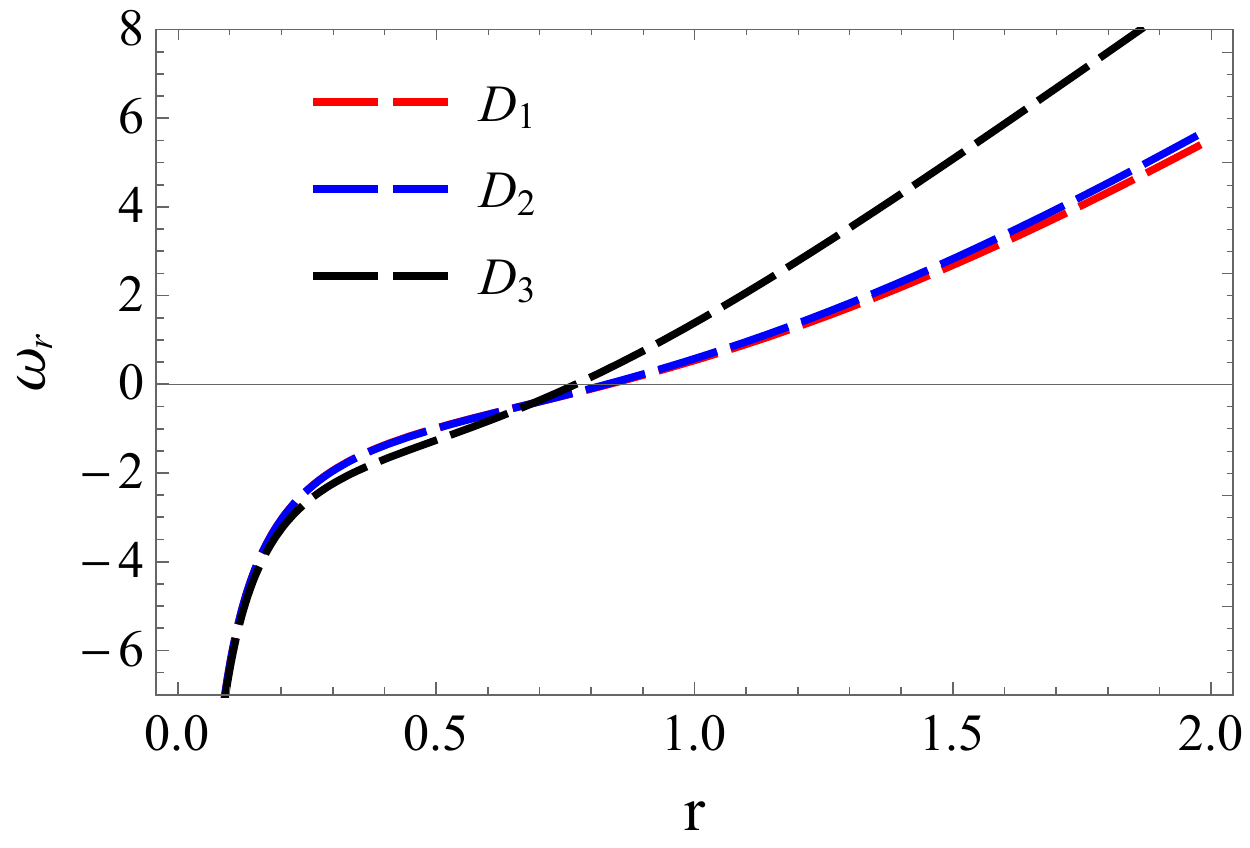}
\caption{The EoS parameter $\omega_r(r)$ against $r$. We use $r_0=1$, $\hbar=1$ and $\beta=0.1$ along with a non-constant redshift
function $\Phi=r_0/r$. }
\end{figure}

\begin{figure}
\includegraphics[width=8.2cm]{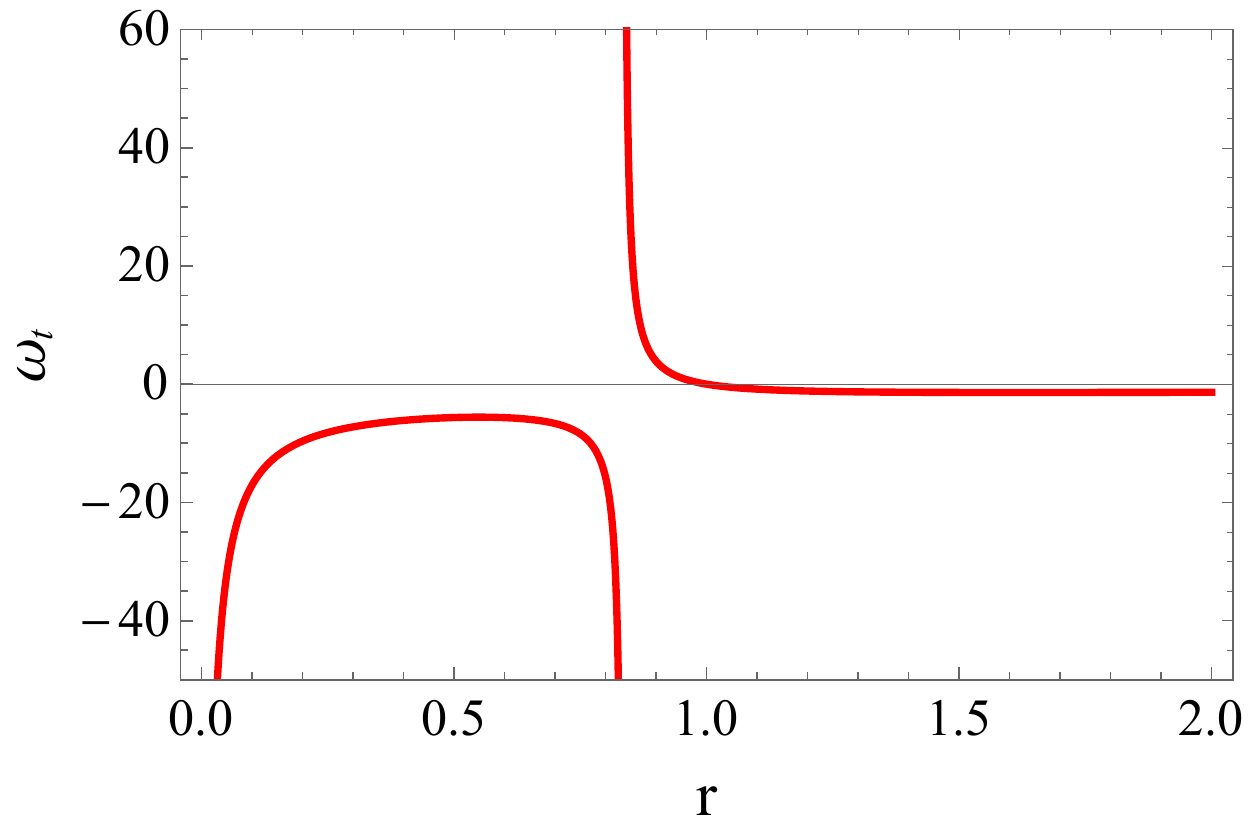}
\caption{The EoS parameter $\omega_t(r)$ for the GUP wormhole with a non-constant redshift
function $\Phi=r_0/r$ as a function of $r$. We use $r_0=1$, $\hbar=1$ and $\beta=0.1$. We consider only the case $D_1$.  }\label{fig4}
\end{figure}

Finally the GUP Casimir wormhole metric can be written as
\begin{equation}
\mathrm{d}s^{2}=-\exp\left({\frac{2 r_0}{r}}\right)\mathrm{d}t^{2}+\frac{\mathrm{d}r^{2}}{1-\frac{b(r)}{r}}+r^{2}\left(
\mathrm{d}\theta ^{2}+\sin ^{2}\theta \mathrm{d}\phi ^{2}\right),  
\end{equation}
with the shape function given by Eq.(\ref{e12}) and satisfies the EoS with the parameter $\omega$ and $n$, given by Eq.(\ref{e27}) and Eq.(\ref{e30}), respectively.

\subsection{Model $\exp(2\Phi(r))=1+\frac{\gamma^2}{r^2}$}
\begin{figure}
\includegraphics[width=8.2cm]{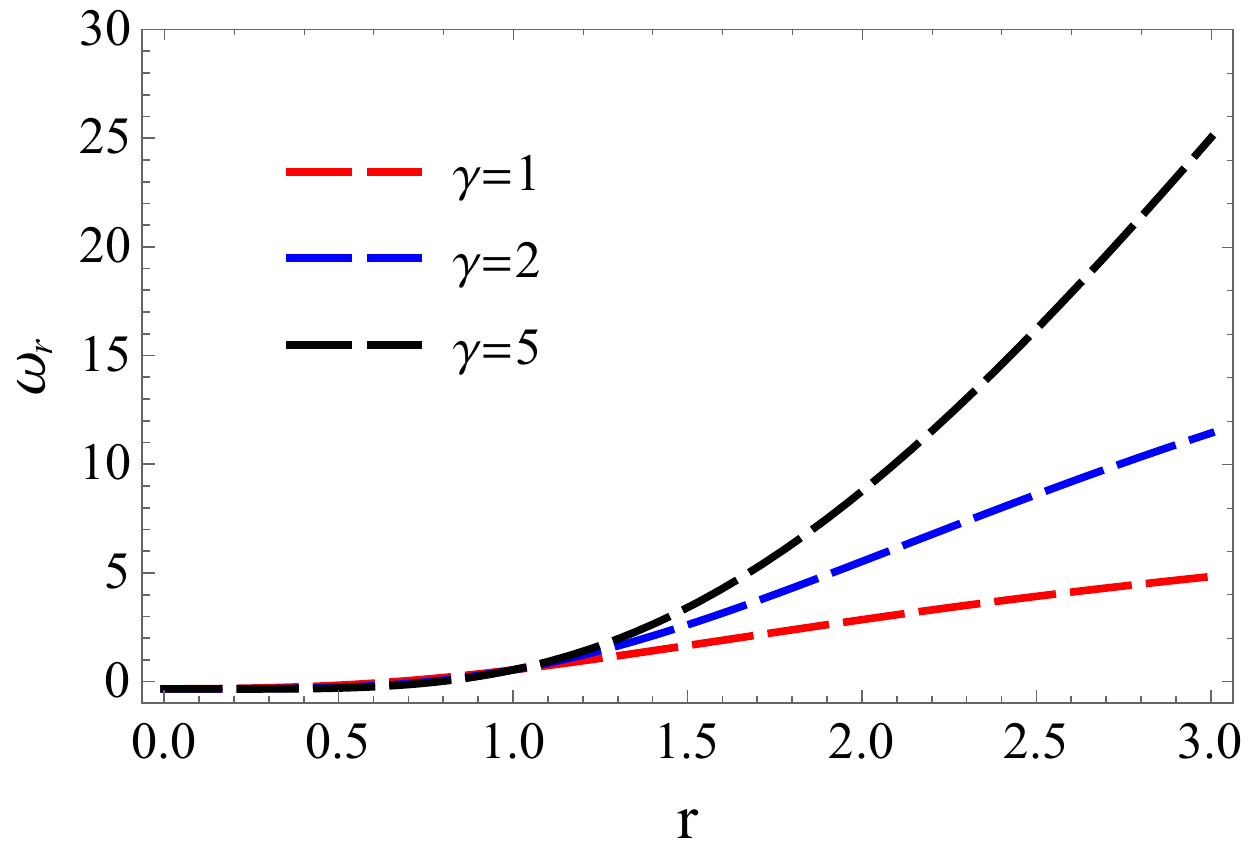}
\caption{The EoS parameter $\omega_r(r)$ against $r$ using the model $\exp(2\Phi(r))=1+\frac{\gamma^2}{r^2}$ for different values of $\gamma$. We use $r_0=1$, $\hbar=1$ and $\beta=0.1$. Here we consider the case $D_1$.  }\label{f6}
\end{figure}

\begin{figure}
\includegraphics[width=8.2cm]{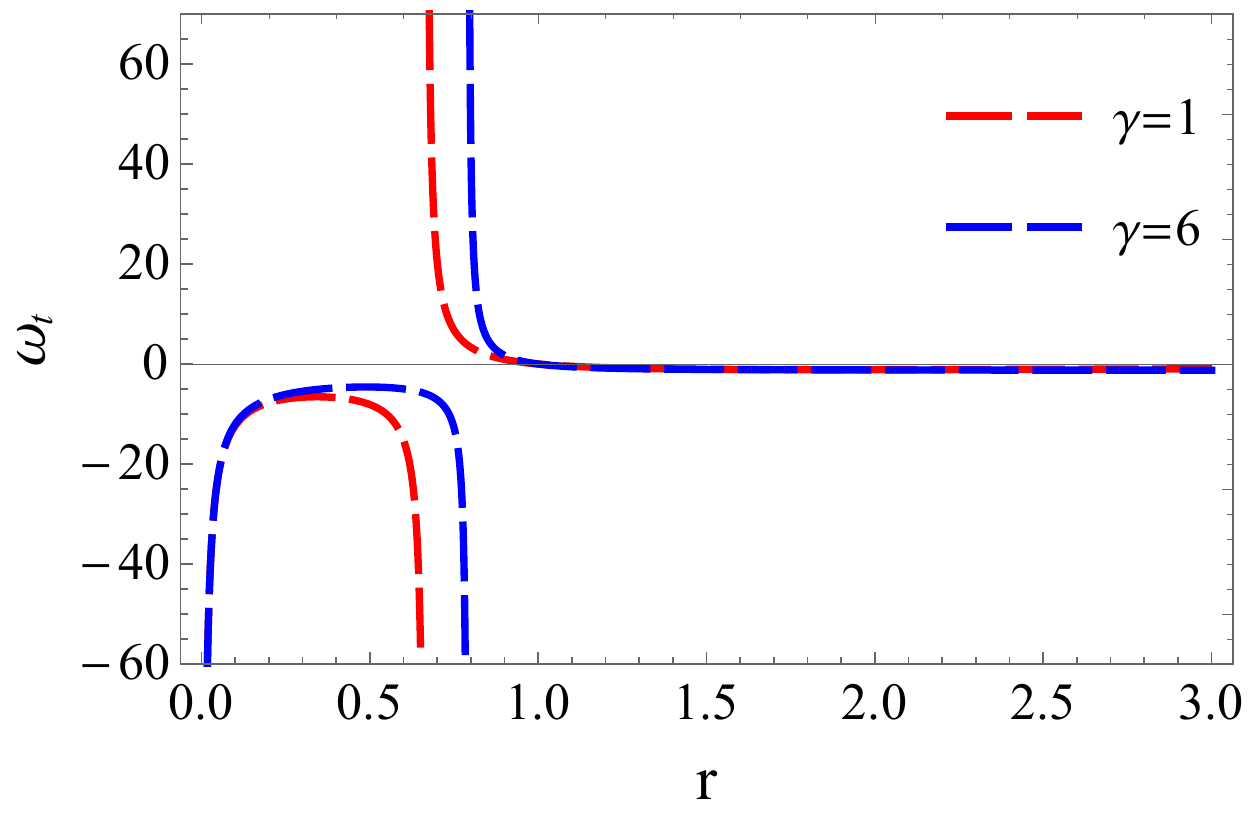}
\caption{The EoS parameter $\omega_t(r)$ against $r$ using the model $\exp(2\Phi(r))=1+\frac{\gamma^2}{r^2}$ and different values of $\gamma$. We use $r_0=1$, $\hbar=1$ and GUP parameter $\beta=0.1$ and $D_1$. }\label{f7}
\end{figure}

Our second example is the following wormhole metric given by
\begin{equation}
\mathrm{d}s^{2}=-\left(1+\frac{\gamma^2}{r^2}\right)\mathrm{d}t^{2}+\frac{\mathrm{d}r^{2}}{1-\frac{b(r)}{r}}+r^{2}\left(
\mathrm{d}\theta ^{2}+\sin ^{2}\theta \mathrm{d}\phi ^{2}\right),
\end{equation}
where $\gamma$ is some positive parameter and $r\geq r_0$. As in the last section, we can assume the EoS $\mathcal{P}_r(r)=\omega_r(r) \rho(r)$ then solve Eq. (\ref{e27}) for the EoS parameter $\omega_r(r)$. Due to the limitation of space, here we can simply skip the full expression for $\omega_r(r)$ and give only the plot for a domain of $\omega_r(r)$ as a function of $r$, illustrated in Fig.\ref{f6}. Finally, we can use the EoS of the form $\mathcal{P}_t(r)=\omega_t(r) \mathcal{P}_r(r)$, and obtain an expression for $\omega_t(r)$. As we already pointed out, we can simply skip the full expression and it is  straightforward  to check the dependence of $\omega_t(r)$ against $r$ given by Fig.\ref{f7}.

\subsection{Isotropic model with $\omega_r(r)=const.$}

From the conservation equation $\nabla_{\mu}T^{\mu\nu}=0$, we can obtain the hydrostatic equation for equilibrium
of the matter sustaining the wormhole
\begin{equation}
    \mathcal{P'}_r(r)=\frac{2(\mathcal{P}_t(r)-\mathcal{P}_r(r))}{r}-(\rho(r)+\mathcal{P}_r(r))\Phi'(r),
\end{equation}
where we have considered a perfect fluid with $\mathcal{P}_t=\mathcal{P}_r$, and assumed the EoS $\mathcal{P}_r(r)=\omega_r \rho(r)$, where $\omega$ is a constant parameter this time. Then it can be reduced to 
\begin{equation}
    \omega_r \rho'(r)=-(1+\omega_r) \rho(r) \,\Phi'(r)
\end{equation}
in which $\rho(r)$ is given by Eq. (\ref{e18}). Solving the last differential equation by setting $\rho(r)\rightarrow -|\rho(r)|$, we obtain the following result
\begin{equation}
    \Phi(r)=C+\frac{\omega_r}{\omega_r+1}\left[\ln\left(\frac{r^6}{r^2+\beta  D_i}\right)\right].\label{iso}
\end{equation}
Absorbing the constant $C$ via the scaling $dt \to C dt$, the wormhole metric element can be written as 
\begin{eqnarray}\notag
\mathrm{d}s^{2}&=&-\left(\frac{r^6}{r^2+\beta  D_i}\right)^{\frac{2}{1+1/\omega}}\mathrm{d}t^{2}+\frac{\mathrm{d}r^{2}}{1-\frac{b(r)}{r}}\\
&+& r^{2}\left(
\mathrm{d}\theta ^{2}+\sin ^{2}\theta \mathrm{d}\phi ^{2}\right),  
\end{eqnarray}
where $r \geq r_0$. It is easy to see that the above solution is finite at the wormhole throat with $r=r_0$, provided $\omega_r \neq -1$. Note that the redshift function $\Phi$ is unbounded for large $r$ as a result one cannot construct asymptotically flat GUP wormholes with isotropic pressures and, in general, such solutions may not be physical. 

\subsection{Anisotropic model with $\omega_r=const.$}

\begin{figure}
\includegraphics[width=8.0 cm]{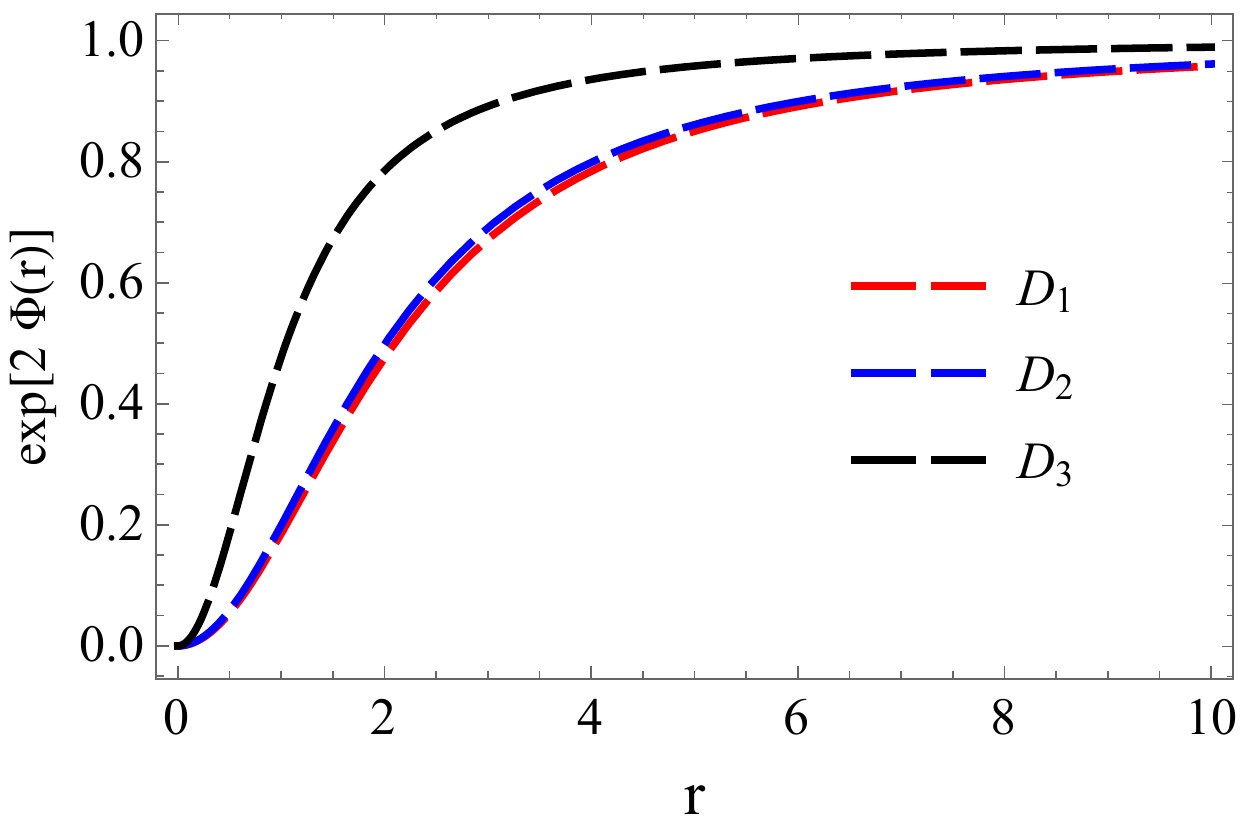}
\caption{We plot $\exp(2\Phi(r))$ for the anisotropic case. We have used $r_0=1$, $\hbar=1$, $\beta=0.1$, $n=-1$ and $\omega=1$. }\label{f88}
\end{figure}

As we have observed, the isotropic model is of very limited physical interest. In this final example we shall elaborate an anisotropic GUP Casimir wormhole spacetime. To do so, can use the following relations $\mathcal{P}_t(r)= n\,\omega_r\, \rho(r)$, and $\mathcal{P}_r(r)=\omega_r \rho(r)$, where $n$ is some constant. We get the following relation
\begin{equation}
    \omega_r\, \rho'(r)=\frac{2\,\omega_r\, \rho(r)\left(n-1\right)}{r}-(1+\omega_r) \rho(r)\,\Phi'(r).
\end{equation}
Solving this equation for the redshift function, we obtain 
\begin{equation}
    \Phi(r)=C+\frac{\omega_r}{\omega_r+1}\left[\ln\left(\frac{r^{2(n+2)}}{r^2+\beta  D_i}\right)   \right].
\end{equation}
We obatain the metric
\begin{eqnarray}\notag
\mathrm{d}s^{2}&=&-\left(\frac{r^{2(n+2)}}{r^2+\beta  D_i}\right)^{\frac{2}{1+1/\omega_r}}\mathrm{d}t^{2}+\frac{\mathrm{d}r^{2}}{1-\frac{b(r)}{r}}\\
&+& r^{2}\left(
\mathrm{d}\theta ^{2}+\sin ^{2}\theta \mathrm{d}\phi ^{2}\right),  
\end{eqnarray}
provided $r \geq r_0$. Notice that we recover an isotropic case (\ref{iso}) here when setting $n=1$ and a singularity at $\omega_r=-1$. In the anisotropic case it is not difficult to show that one can construct asymptotically flat spacetime. Setting $n=-1$ and  $\omega_r \neq -1$, the above metric reduces to 
\begin{eqnarray}\notag
\mathrm{d}s^{2}&=&-\left(\frac{1}{1+\frac{\beta  D_i}{r^2}}\right)^{\frac{2}{1+1/\omega_r}}\mathrm{d}t^{2}+\frac{\mathrm{d}r^{2}}{1-\frac{b(r)}{r}}\\
&+& r^{2}\left(
\mathrm{d}\theta ^{2}+\sin ^{2}\theta \mathrm{d}\phi ^{2}\right),  
\end{eqnarray}
which is asymptotically flat spacetime. In fact, it is easy to check that the case $n=-1$ gives the only asymptotically 
at flat solution. As we can see from Fig.\ref{f88}, in the limit $r \to \infty$ we obtain $\exp(2\Phi(r))=1$, as was expected.

\section{Embedding Diagram}
\begin{figure*}
\includegraphics[width=6.0 cm]{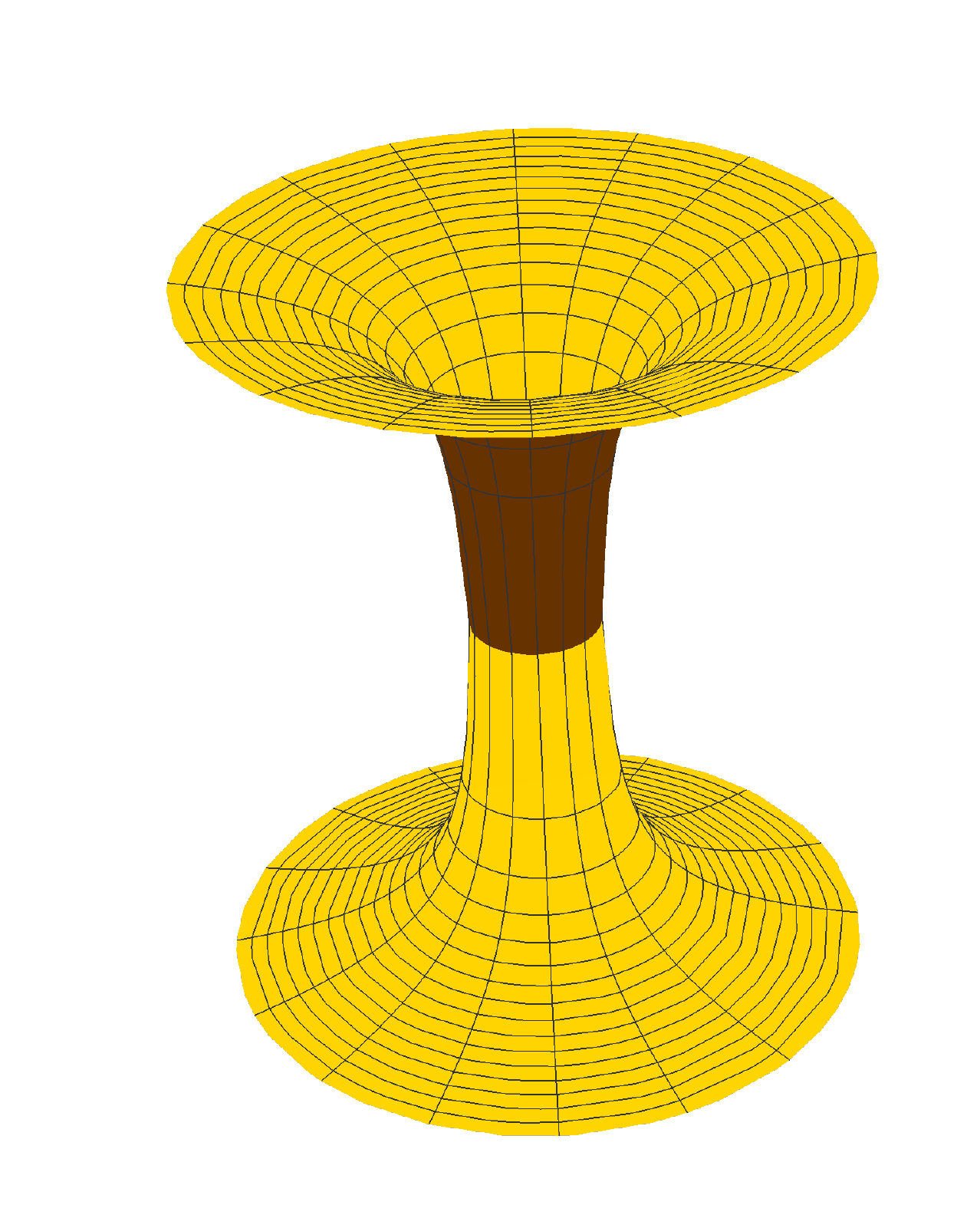}\hspace{1.5 cm}
\includegraphics[width=6.5 cm]{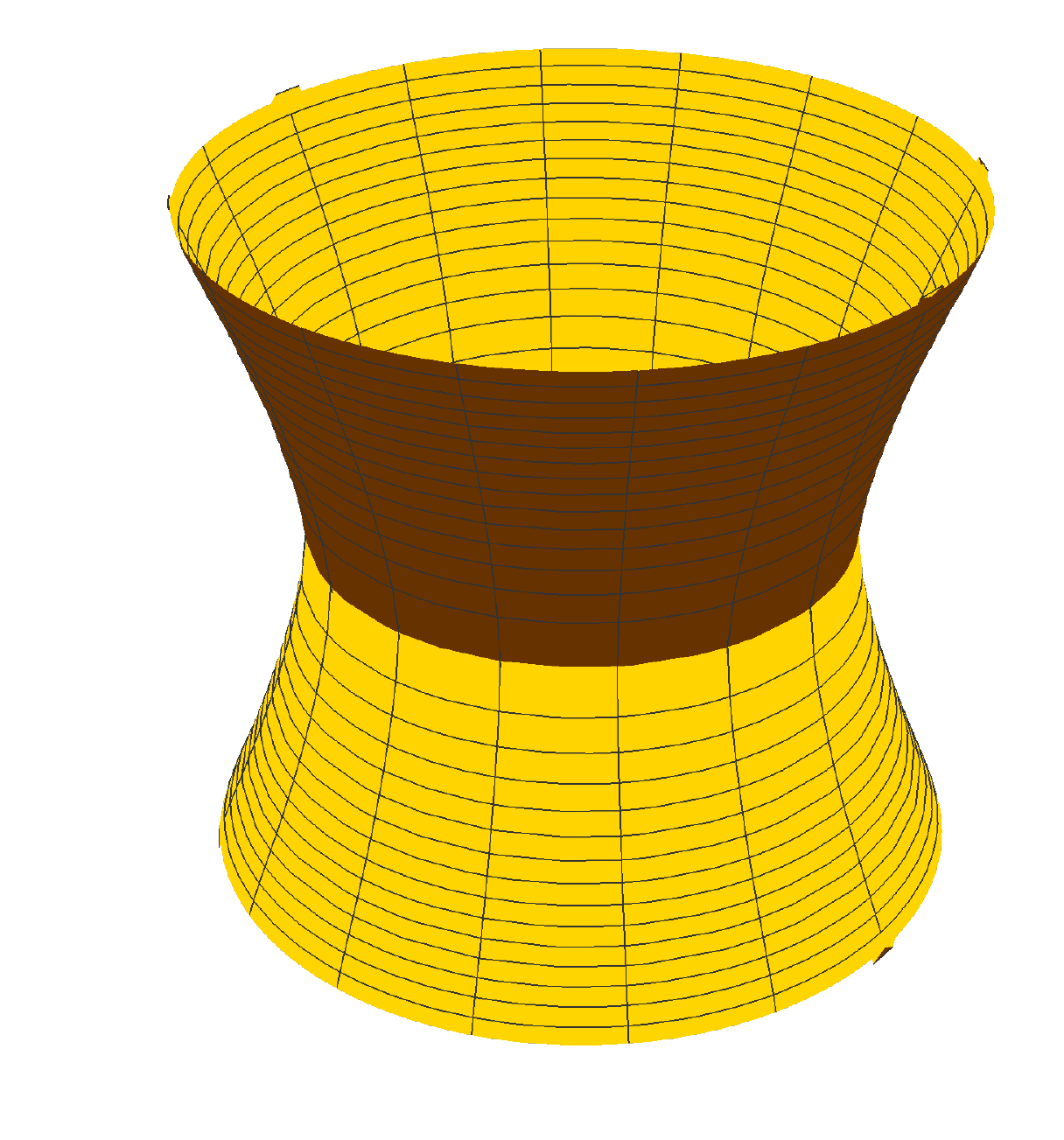}
\caption{The GUP Corrected Casimir wormhole embedded
in a three-dimensional Euclidean space. Left panel: We have used $r_0=1$, $\hbar=1$, $\beta=0.06$. Right panel: We have used $r_0=1$, $\hbar=1$, $\beta=0.18$. In both plots we have used $D_1$. }\label{f8}
\end{figure*}

In this section we discuss the embedding diagrams to represent the GUP corrected Casimir wormhole by considering an equatorial slice $\theta=\pi/2$ at some fix moment in time $t=constant$. The metric can be written as 
\begin{equation}
    ds^2=\frac{dr^2}{1-\frac{b(r)}{r}}+r^2d\phi^2.\label{emb}
\end{equation}
We embed the metric (\ref{emb}) into three-dimensional Euclidean space to visualize this slice and
the spacetime can be written in cylindrical coordinates as
\begin{equation}
    ds^2=dz^2+dr^2+r^2d\phi^2.
\end{equation}
From the last two equations we find that 
\begin{equation}
    \frac{dz}{dr}=\pm \sqrt{\frac{r}{r-b(r)}-1}.
\end{equation}
where $b(r)$ is given by Eq. (\ref{s22}). Note that the integration of the last expression cannot
be accomplished analytically. Invoking numerical techniques allows us to illustrate the
wormhole shape given in Fig.\ref{f8}. From Fig. \ref{f8} we observe the effect of GUP parameter on the wormhole geometry.

\section{ADM Mass of GUP wormhole}
Now let us compute the ADM mass for GUP Casimir wormhole.  We consider the asymptotic flat 
spacetime
\begin{eqnarray}
ds^2_{\Sigma} = \psi(r)dr^2+r^2 \chi(r)\left(d\theta^2+\sin^2\theta d\phi^2\right),
\end{eqnarray}
where we have identified
\begin{equation}
    \psi(r)=\frac{1}{1-\frac{b(r)}{r}},\quad \text{and} \quad \chi(r)=1.
\end{equation}
In order to compute the ADM mass, we use the approach followed the following relation (see, \cite{Shaikh:2018kfv}):
\begin{equation}\label{for}
    M_{ADM}=\lim_{r\to \infty} \frac{1}{2}\left[-r^2 \chi'+r(\psi -\chi) \right].
\end{equation}
On substituting the values in (\ref{for}) and after computing the limit we get the ADM mass for the 
wormhole,
\begin{equation}\label{ADM}
    M_{ADM}=r_0-\frac{\pi^3 }{90 r_0}-\frac{\beta D_{i} \pi^3 }{270 r_0^3}.
\end{equation}
Note that this is the mass of the  wormhole as seen by an observer located at the 
asymptotic spatial infinity. It is observed that the GUP effect decreases  the ADM mass. Notice that the ADM mass (\ref{ADM}) consists of three terms: the geometric term $r_0$ given by the first term, a semiclassical quantum effect of the spacetime given by the second term, and finally the GUP effect given by the third term. Related to the GUP parameter, let us point our that in Ref. \cite{Das:2008kaa} authors have speculated about the possibility to predict upper bounds on the quantum gravity parameter in the GUP, compatible with experiments at the electroweak scale.

\section{Energy Conditions}
\label{Ener}
Given the redshift function and the shape function, we can compute the energy-momentum components. In
particular for the radial component we find
\begin{equation}
    \mathcal{P}_r=\frac{\beta D_{i}  \pi^3  (r^4-2 r^3 r_0-r r_0^3+2 r_0^4)+\mathcal{F} }{2160 r^7 r_0^3 \pi},
\end{equation}
where $\mathcal{F} $ is given by Eq. (\ref{e31}). On the other hand for the tangential component of the pressure we find the following result
\begin{equation}
    \mathcal{P}_t=\frac{(r-r_0) \left[\beta D_{i} \pi^3 (r^2+r r_0+r_0^2)\mathcal{H}+3r_0^2 r^2 \mathcal{G}\right]}{4320 r_0^3 \pi r^8},\label{fig5}
\end{equation}
in which $\mathcal{H}$ and $\mathcal{G}$ are given by Eq. (\ref{e31}) and (\ref{e32}), respectively.
\begin{figure}
\includegraphics[width=8cm]{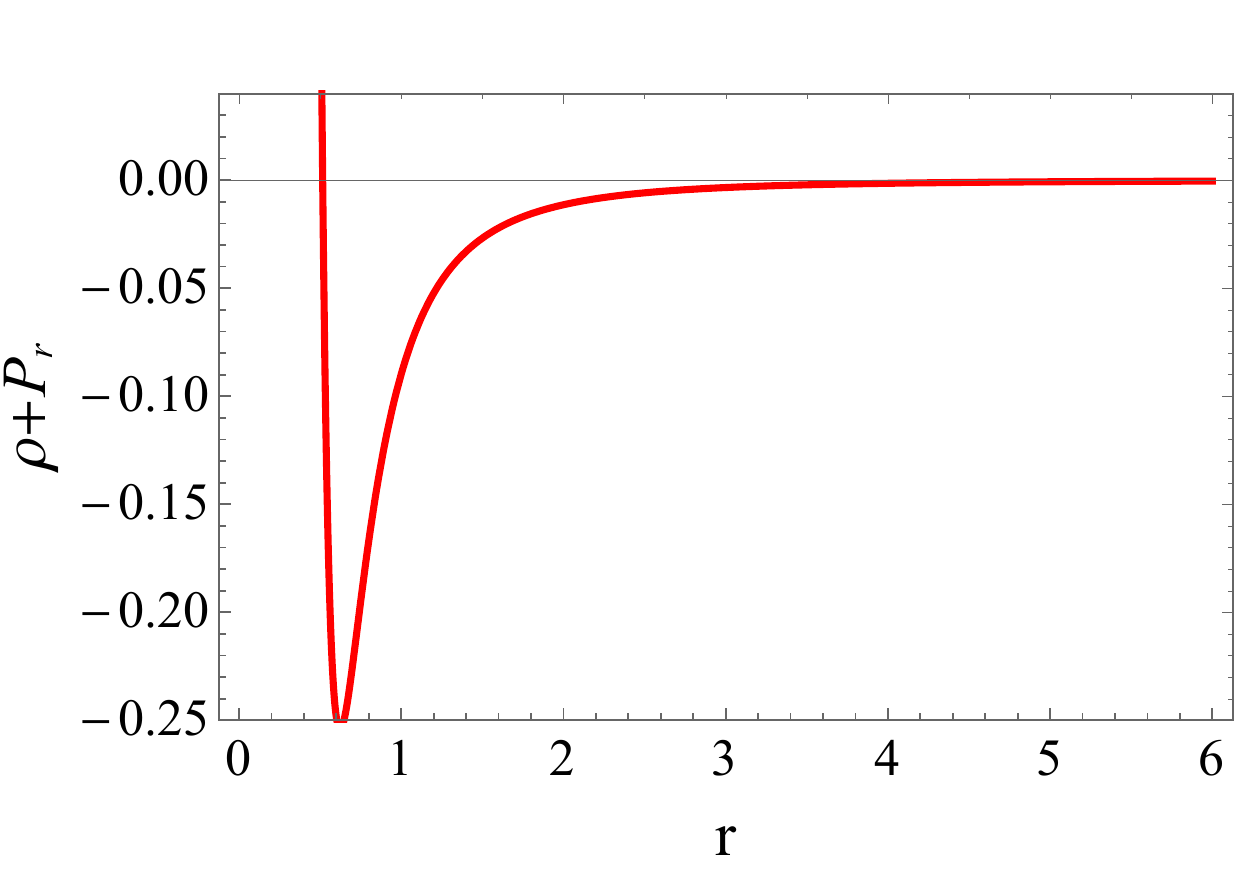}
\caption{ The variation of $\rho+\mathcal{P}_r$ as a function of $r$ using $\Phi=r_0/r$. We use $r_0=1$, $\hbar=1$, $\beta=0.1$ and $D_1$.  }
\end{figure}
\begin{figure}
\includegraphics[width=8cm]{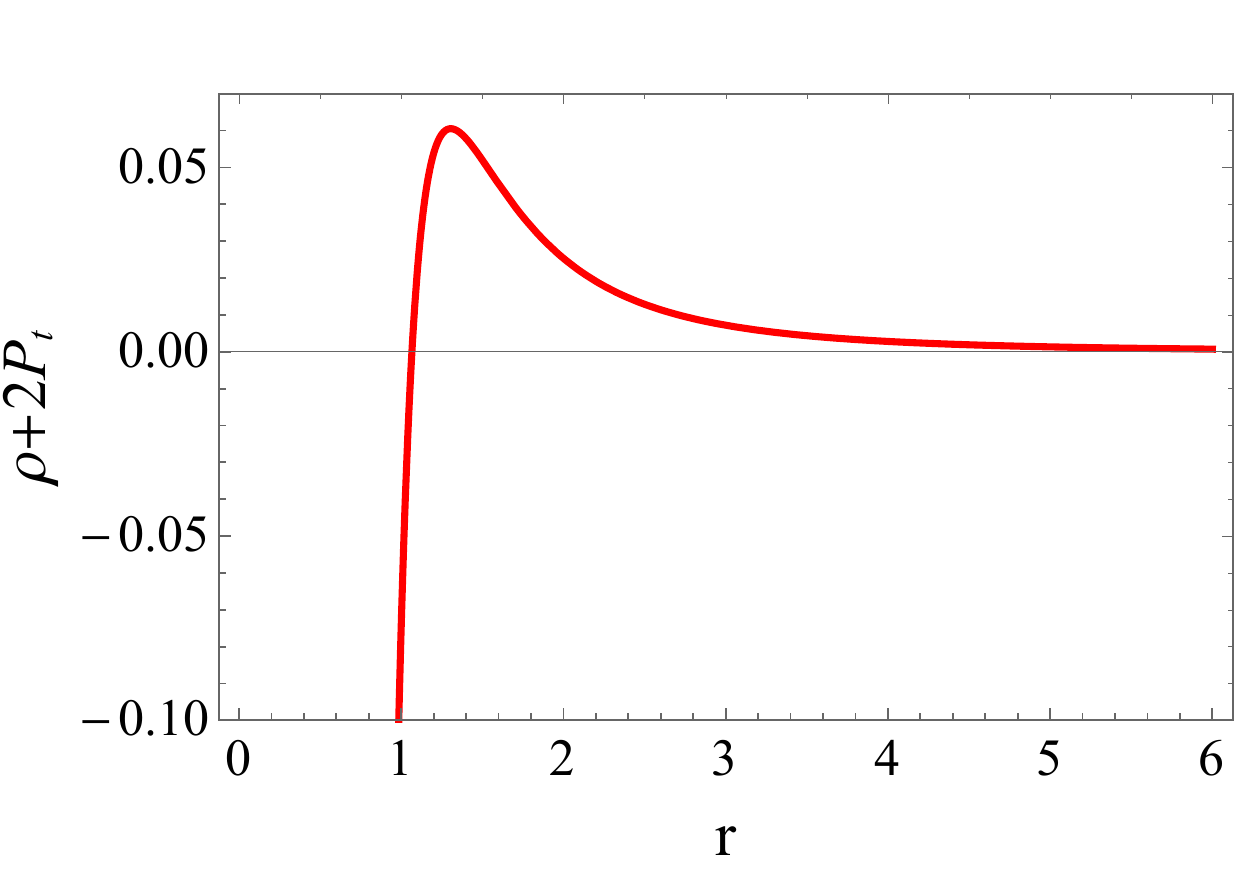}
\caption{ The variation of $\rho+2\mathcal{P}_t$ against $r$ and $\Phi=r_0/r$. We use $r_0=1$, $\hbar=1$, $\beta=0.1$ and $D_1$.  }\label{fig6}
\end{figure}
\begin{figure}
\includegraphics[width=8cm]{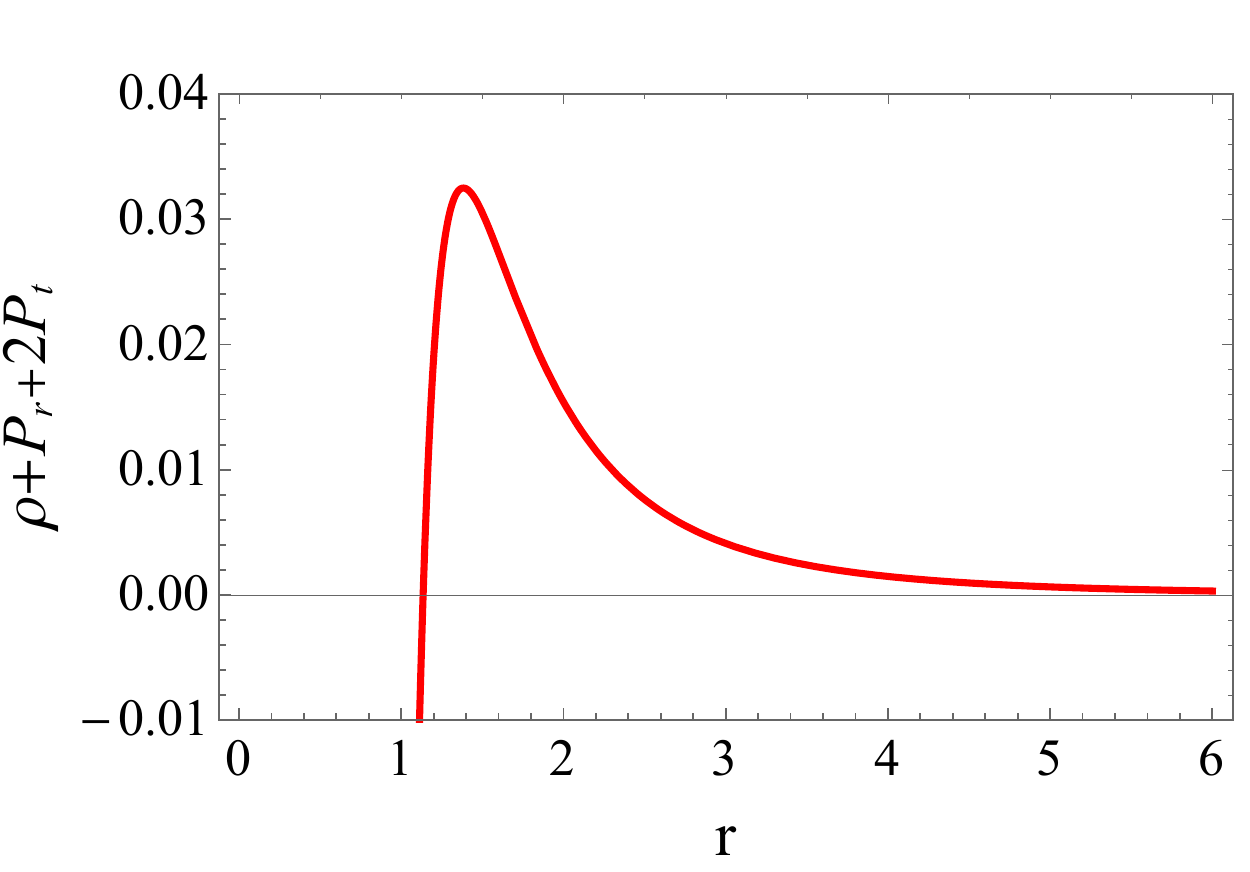}
\caption{ The variation of $\rho+\mathcal{P}_r+2 \mathcal{P}_t$ against $r$ and $\Phi=r_0/r$. We use $r_0=1$, $\hbar=1$, $\beta=0.1$ and $D_1$.}\label{fig7}
\end{figure}

With these results we can continue our our discussion on the issue of energy conditions and make some
regional plots to check the validity of all energy conditions. In particular we recall that the WEC is
defined by $T_{\mu \nu }U^{\mu }U^{\nu }\geq 0$ i.e.,
\begin{equation}
\rho (r)+\mathcal{P}_{r}(r)\geq 0,
\end{equation}
where $T_{\mu \nu }$ is the energy momentum tensor and $U^{\mu }$ denotes the timelike vector. In other words, the
local energy density is positive and it gives rise to the continuity of NEC,
which is defined by $T_{\mu \nu }k^{\mu }k^{\nu }\geq 0$ i.e., 
\begin{equation}
\rho (r)+\mathcal{P}_{r}(r)\geq 0, 
\end{equation}
where $k^{\mu }$ is a null vector. On the other hand the strong energy condition (SEC) stipulates that
\begin{equation}
\rho (r)+2\mathcal{P}_t(r)\geq 0,
\end{equation}
and
\begin{equation}
\rho (r)+\mathcal{P}_{r}(r)+2\mathcal{P}_t(r)\geq 0.
\end{equation}

\begin{figure}
\includegraphics[width=8cm]{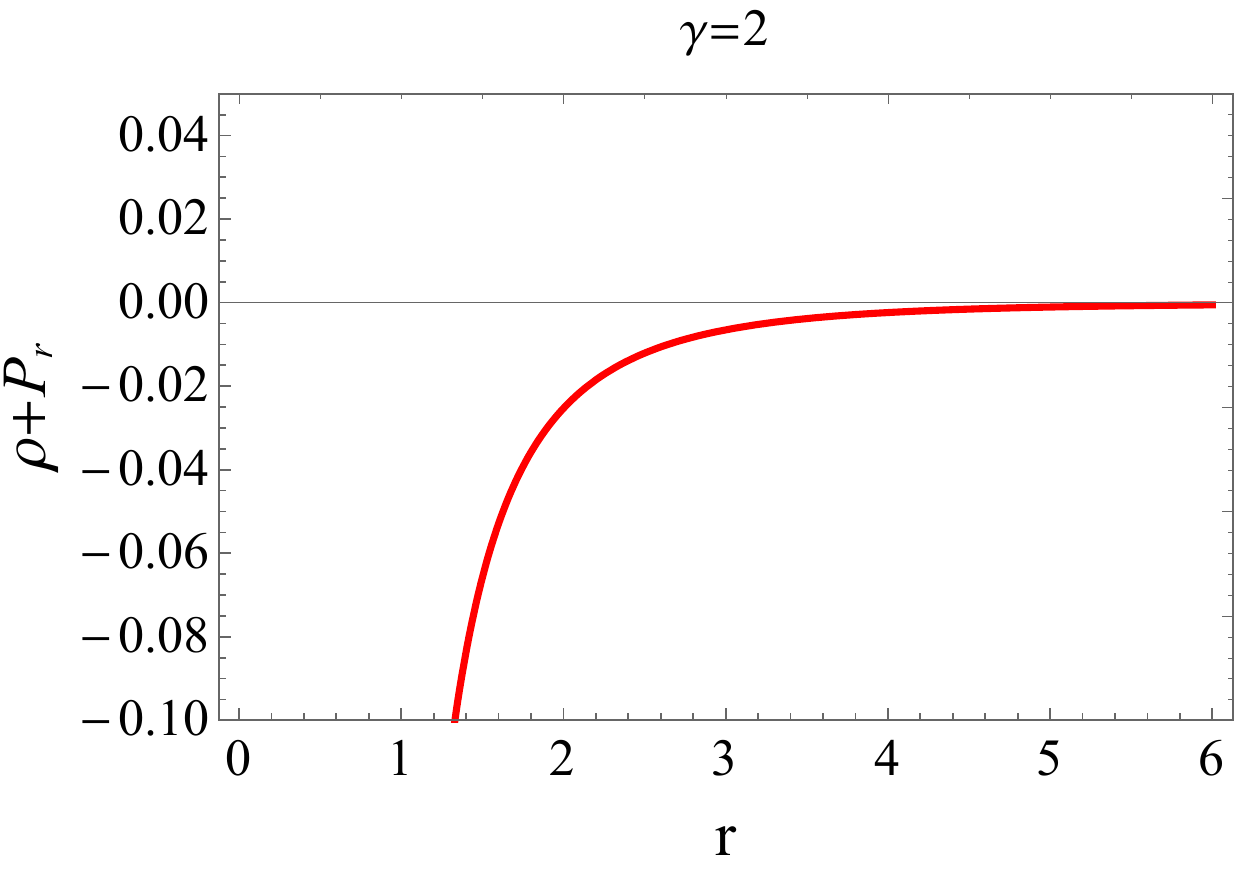}
\caption{ The variation of $\rho+\mathcal{P}_r$ as a function of $r$ using $\exp(2\Phi(r))=1+\frac{\gamma^2}{r^2}$. We use $r_0=1$, $\hbar=1$ and $\beta=0.1$.  }\label{fig11}
\end{figure}

\begin{figure}
\includegraphics[width=8cm]{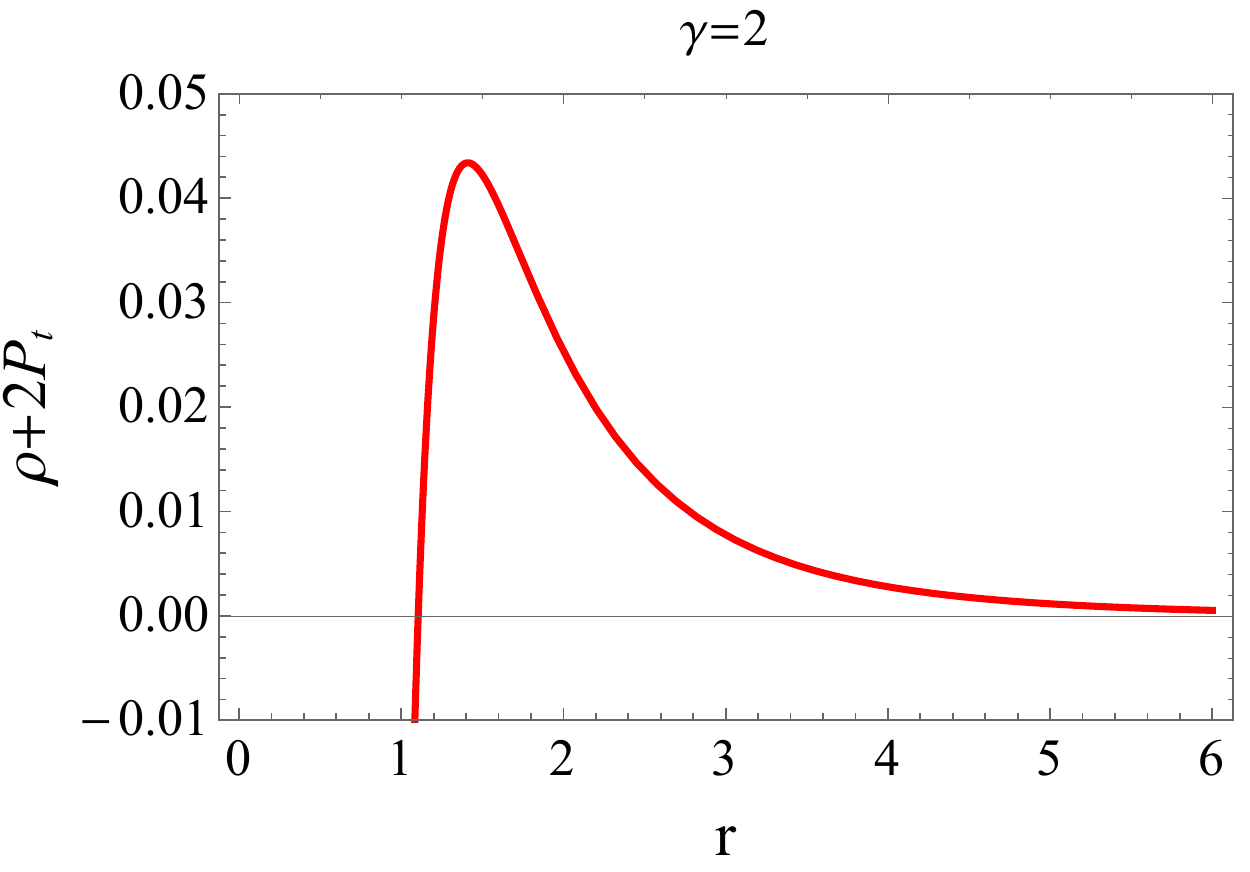}
\caption{ The variation of $\rho+2\mathcal{P}_t$ as a function of $r$ using $\exp(2\Phi(r))=1+\frac{\gamma^2}{r^2}$. We use $r_0=1$, $\hbar=1$ and $\beta=0.1$.  }\label{fig12}
\end{figure}

\begin{figure}
\includegraphics[width=8cm]{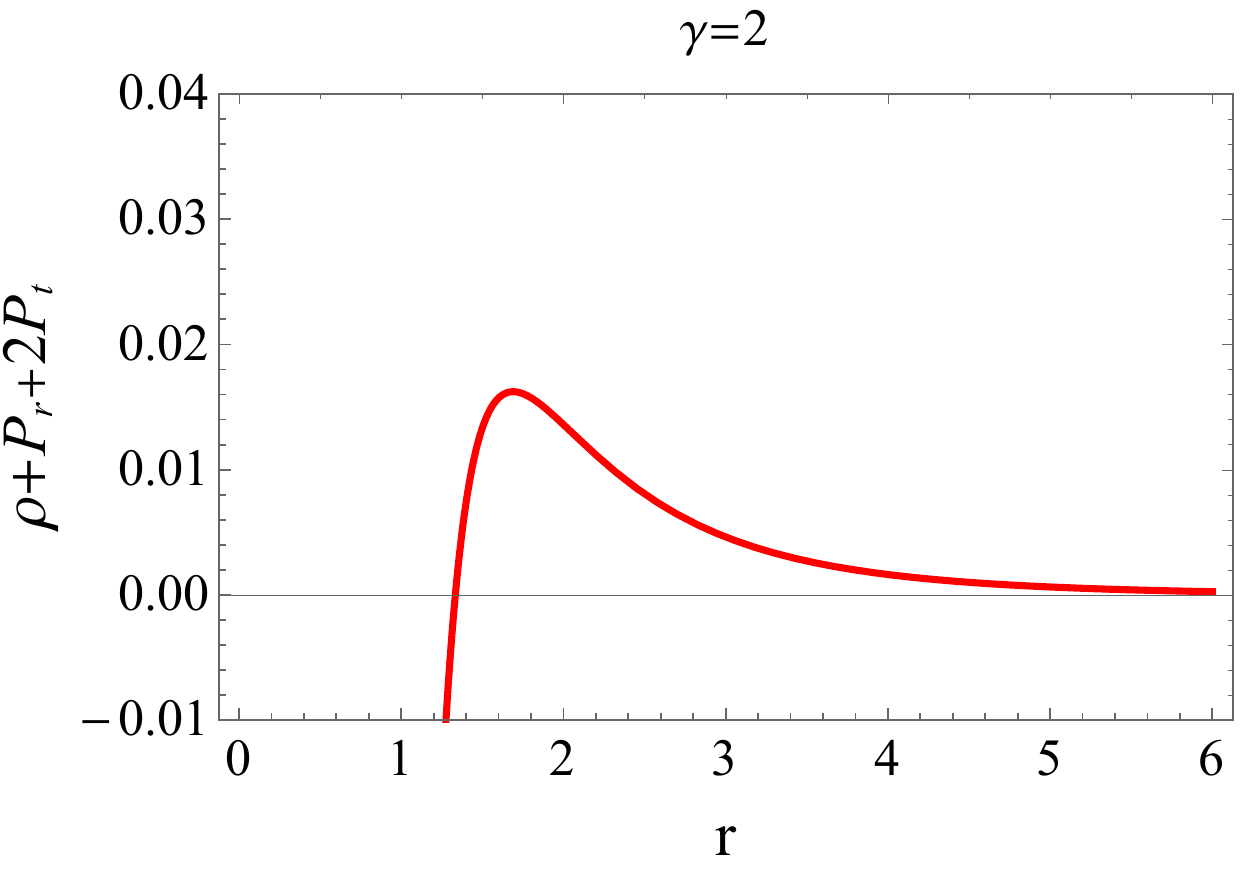}
\caption{ The variation of $\rho+\mathcal{P}_r+2\mathit{P}_t$ as a function of $r$ using $\exp(2\Phi(r))=1+\frac{\gamma^2}{r^2}$. We use $r_0=1$, $\hbar=1$ and $\beta=0.1$.  }\label{figure13}
\end{figure}

We see from Figs.(\ref{fig4}-\ref{fig7}), and similarly Figs.(\ref{fig11}-\ref{figure13}), NEC, WEC, and SEC, are not satisfied at the wormhole throat $r=r_0$.  In fact one can check numerically that in all plots at the wormhole throat $r=r_0$, we have $\left(\rho+\mathcal{P}_r\right)\vert_{r_0=1}<0$, along with $\left(\rho+\mathcal{P}_r+2\mathcal{P}_t\right)\vert_{r_0=1}<0$, by small values.

However, from the quantum field theory it is known that quantum fluctuations violate most energy conditions without any restrictions and this opens the possibility that quantum fluctuations may play an important role in the wormhole stability. For instance, one can
examine the consequences of the constraint imposed by a Quantum Weak Energy Condition (QWEC) given by \cite{Garattini:2019ivd}
\begin{equation}
\rho (r)+\mathcal{P}_{r}(r)<f(r), \,\,\,f(r)>0,
\end{equation}
where $r \in[r_0,\infty)$. Thus such small violations of energy conditions due to  the quantum fluctuations are possible in quantum field theory. 

\section{Amount of exotic matter}
\label{amont}
In this section we shall briefly discuss  the ``volume integral quantifier,'' which basically quantifies the amount of exotic matter required for wormhole maintenance. This quantity is related only to $\rho$  and $\mathcal{P}_r$, not to the transverse components, and is defined in terms of the following definite integral
\begin{eqnarray}
\mathcal{I}_V=\oint [\rho+\mathcal{P}_r]~\mathrm{d}V=2 \int_{r_0}^{\infty} \left(  \rho+\mathcal{P}_r\right)~\mathrm{d}V,  
\end{eqnarray}
which can be written also as
\begin{eqnarray}
\mathcal{I}_V =8 \pi \int_{r_0}^{\infty} \left(  \rho+\mathcal{P}_r  \right)r^2  dr.
\end{eqnarray}
As we already pointed out, the value of this volume-integral encodes information about the ``total amount"
of exotic matter in the spacetime, and we are going to evaluate this integral for 
our shape function $b(r)$. It is convenient to introduce a cut off such that the wormhole extends form $r_0$ to a radius situated at $`a'$ and then  we get the very simple result
\begin{equation}
\mathcal{I}_V=   8 \pi \int_{r_0}^{a} \left(  \rho+\mathcal{P}_r  \right)r^2  dr.
\end{equation}

\begin{figure}
\includegraphics[width=8.2cm]{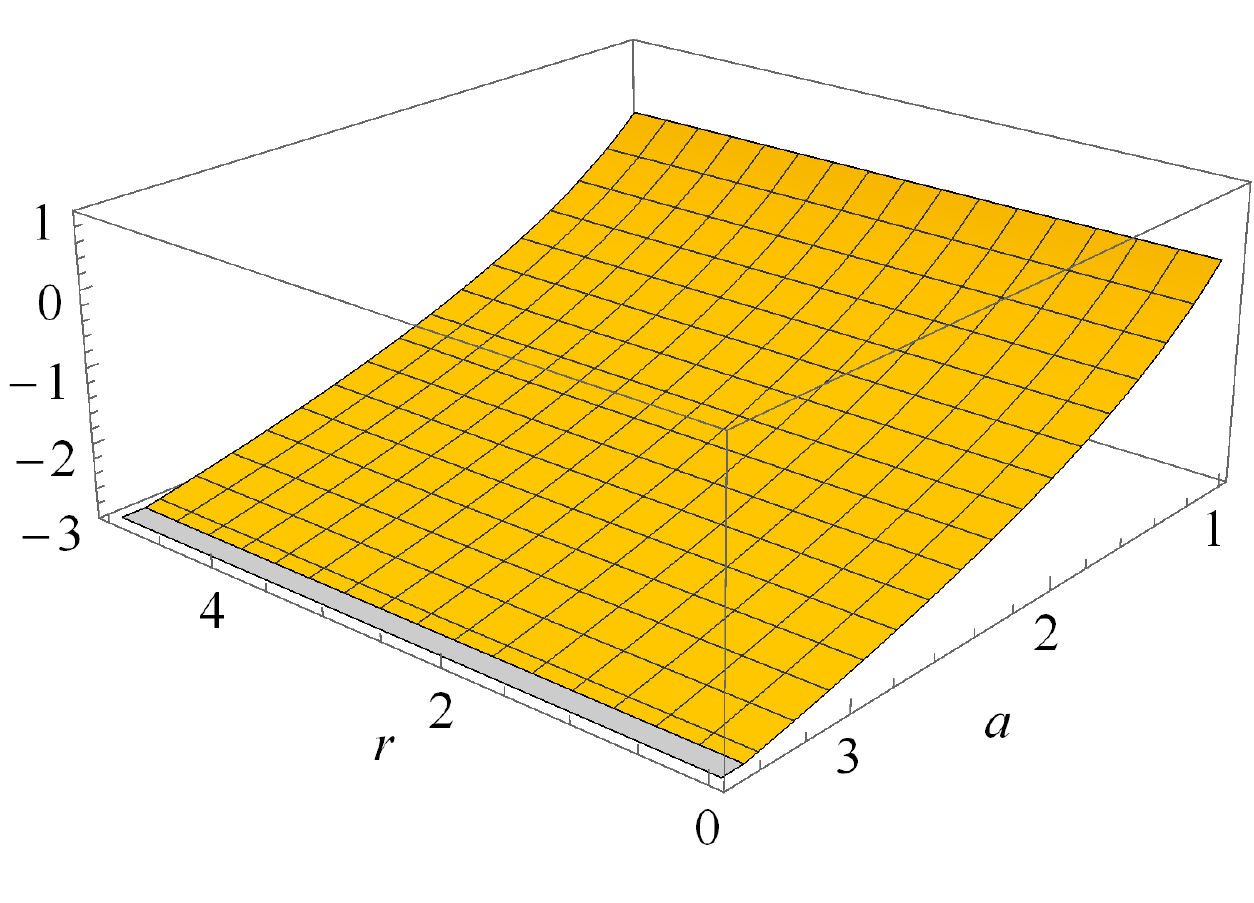}
\caption{ The variation of $\mathcal{I}_V$ against $r$ and $a$ of the case $\Phi=r_0/r$. We use $r_0=1$, $\hbar=1$ and $\beta=0.1$.}\label{fig166}
\end{figure}

\begin{figure}
\includegraphics[width=8.2cm]{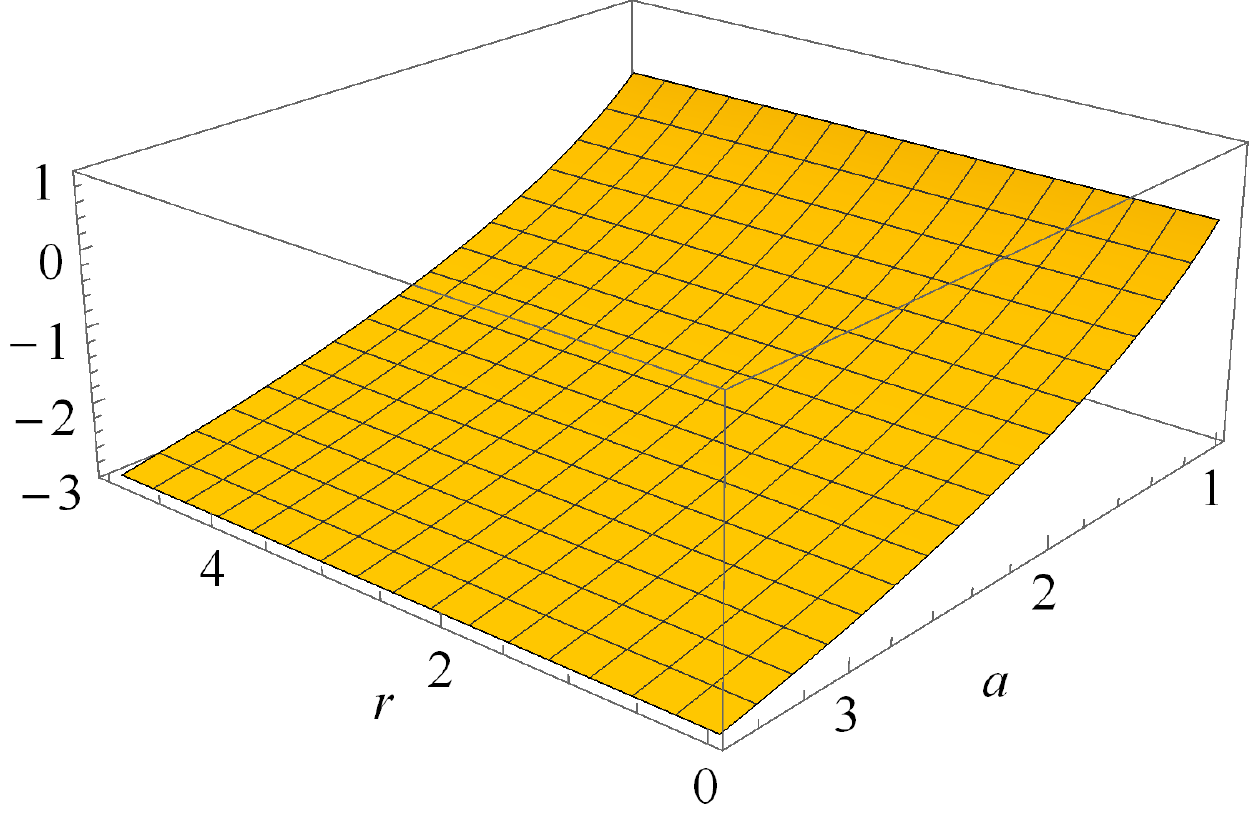}
\caption{ The variation of $\mathcal{I}_V$ against $r$ and $a$ of the case $\exp(2\Phi(r))=1+\frac{\gamma^2}{r^2}$. We use $r_0=1$, $\gamma=2$, $\hbar=1$ and $\beta=0.1$.  }\label{fig17}
\end{figure}

In the special case $a \rightarrow r_0$, we should find $\int{(\rho+\mathcal{P}_r)} \rightarrow 0$. In the specific case having $\Phi=r_0/r$, the Casimir wormhole is supported by arbitrarily small quantities of exotic matter. Evaluating the above integral we find that
\begin{equation}
    \mathcal{I}_V=\frac{\beta D_{i}  \pi^3 (a-r_0)\mathcal{M}+18 a^2 r_0^2 \mathcal{N}}{1620 a^4 r_0^3},
\end{equation}
where
\begin{equation}
    \mathcal{M}=6 a^4 \ln(\frac{a}{r_0})-17 a^4+12a^3 r_0+8 a r_0^3-3 r_0^4,
\end{equation}
and
\begin{eqnarray}
    \mathcal{N}&=&\ln(\frac{a}{r_0})(\pi^3-270r^2)\nonumber\\&&-3(a-r_0)[(\pi^3-60 r_0^2)a-\pi^3 r_0].
\end{eqnarray}
From Fig.(\ref{fig166}), we observe that the quantity $\mathcal{I}_V$ is negative, i.e., $\mathcal{I}_V<0$. On the other hand we can also use the redshift $\exp(2\Phi(r))=1+\frac{\gamma^2}{r^2}$ to obtain an expression for the amount of exotic matter. Due to the limitation of space, we are going to skip the full expression for $\mathcal{I}_V$ and give only the dependence of $\mathcal{I}_V$ against $r$ and $a$, given by Fig.(\ref{fig17}).  Hence it demonstrates the existence of spacetime geometries containing traversable wormholes that are supported by arbitrarily small quantities of ``exotic matter''. Such small violations of this quantity can be linked to the quantum fluctuations. We leave this interesting topic for further investigation. 
\section{Light deflection by GUP Casimir wormhole}
\label{chvi}
\subsection{Case with $\Phi(r)=const$.}
In this section we shall proceed to explore the gravitational lensing effect in the spacetime of the GUP Casimir wormhole with $\Phi(r)=const$. The optical metric of GUP wormhole, in the equatorial plane, is simply found by letting $\mathrm{d}s^2=0$, yielding
\begin{eqnarray}\notag
dt^2&=&\frac{dr^2}{1-\frac{r_0}{r}-\frac{\pi^3}{90 r}\left(\frac{1}{r}-\frac{1}{r_0}\right)-\frac{\pi^3 D_{i} \beta}{270 r} \left(\frac{1}{r^3}-\frac{1}{r_0^3}\right)}\\
&+&r^2d\phi^2.
\end{eqnarray}
In the present paper, we are going to use a recent geometric method based on the Gauss-Bonnet theorem (GBT) to calculate the deflection angle. Let $\mathcal{A}_{R}$  be a non-singular domain (or a region outside the light ray) with boundaries $\partial 
\mathcal{A}_{R}=\gamma_{g^{(op)}}\cup C_{R}$, of an oriented two-dimensional surface $S$ with the optical metric $g^{(op)}$. Furthermore let $K$ and $\kappa $ be the Gaussian optical
curvature and the geodesic curvature, respectively. Then, the GBT can be stated as follows \cite{Jusufi:2018waj}
\begin{equation}
\iint\limits_{\mathcal{A}_{R}}K\,\mathrm{d}S+\oint\limits_{\partial \mathcal{%
A}_{R}}\kappa \,\mathrm{d}t+\sum_{k}\theta _{k}=2\pi \chi (\mathcal{A}_{R}).
\label{10}
\end{equation}
In which $\mathrm{d}S$ is the optical surface element, $\theta _{k}$ gives the exterior angle at the $k^{th}$ vertex. Basically the GBT provides a relation between the geometry and the topology of the spacetime. By construction, we need to choose the domain of integration to be outside of the light
ray in the $(r,\phi)$ optical plane. Moreover this domain can be thought to have the topology of disc having the Euler characteristic number $\chi (\mathcal{A}_{R})=1$. Next, let us introduce a smooth curve defined as $\gamma:=\{t\}\to \mathcal{A}_{R}$,  with the geodesic curvature defined by the following relation
\begin{equation}
\kappa =g^{(op)}\,\left( \nabla _{\dot{\gamma}}\dot{\gamma},\ddot{\gamma}%
\right),  
\end{equation}%
 along with the  unit speed condition $g^{(op)}(\dot{\gamma},\dot{\gamma})=1$, and $\ddot{\gamma}$ being the unit acceleration vector. Now if we consider a very large, but finite radial distance $l\equiv R\rightarrow \infty $, such that the two jump angles (at the source $\mathcal{S}$, and observer $\mathcal{O})
$, yields  $\theta _{\mathit{O}}+\theta _{\mathit{S}}\rightarrow \pi $. Note that, by definition, the geodesic curvature for the light ray (geodesics) $\gamma_{g^{(op)}}$ vanishes, i.e. $\kappa (\gamma_{g^{(op)}})=0$. One should only compute the contribution to the curve $C_{R}$. That being said, from the GBT we find
\begin{equation}
\lim_{R\rightarrow \infty }\int_{0}^{\pi+\hat{\alpha}}\left[\kappa \frac{d t}{d \phi}\right]_{C_R} d \phi=\pi-\lim_{R\rightarrow \infty }\iint\limits_{\mathcal{A}_{R}}K\,\mathrm{d}S
\end{equation}
The geodesic curvature for the curve $C_{R}$ located at a coordinate distance $R$ from the coordinate system chosen at the ringhole center can be calculated via the relation 
\begin{equation}
\kappa (C_{R})=|\nabla _{\dot{C}_{R}}\dot{C}_{R}|.
\end{equation}
With the help of the unit speed condition, one can show that the asymptotically Euclidean condition is satisfied:
\begin{eqnarray}
\lim_{R\rightarrow \infty } \left[\kappa \frac{\mathrm{d}t}{\mathrm{d}\phi}\right]_{C_{R}}=1.
\end{eqnarray}
From the GBT it is not difficult to solve for the deflection angle which gives
\begin{equation}
\hat{\alpha}=-\int\limits_{0}^{\pi }\int\limits_{r=\frac{b}{\sin \phi}%
}^{\infty } K \mathrm{d}S. 
\end{equation}
where an equation for the light ray is $r(\phi)=\mathsf{b}/\sin \phi $.  The Gaussian optical curvature takes the form: 
\begin{equation}
    K=\frac{3 r_0^2 r^2 [(\pi^3-90 r_0^2)r-2 \pi^3  r_0]+ \beta D_{i} \pi^3  (r^3-4 r_0^3) }{540 r^6 r_0^3}.
\end{equation}
Approximating this expression in leading order, the deflection angle reads
\begin{equation}
\hat{\alpha}=-\int\limits_{0}^{\pi }\int\limits_{\frac{\mathsf{b}}{\sin \phi }%
}^{\infty }\left[  -\frac{ r_0}{2r^3}+\frac{ \pi^3 (r-2r_0)}{180 r^4 r_0}\right]r dr d\phi. 
\end{equation}
Solving this integral, we find the following solution
\begin{equation}
\hat{\alpha}\simeq \frac{r_0}{\mathsf{b}}-\frac{\pi^3 }{90 r_0 \mathsf{b}}\left(1-\frac{\pi r_0}{4\,\mathsf{b}}\right).\label{alga}
\end{equation}
We see that the first term is due to the wormhole geometry, while the second term is related to the  semiclassical quantum effects of the spacetime.

\subsection{Case with $\Phi(r)=r_0/r$.}
In this case, the optical metric in the equatorial plane takes the form 
\begin{eqnarray}\notag
dt^2&=&\frac{\exp[-{\frac{2r_0}{r}}] dr^2}{1-\frac{r_0}{r}-\frac{\pi^3 }{90 r}\left(\frac{1}{r}-\frac{1}{r_0}\right)-\frac{\pi^3 \hbar^3 D_{i} \beta}{270 r} \left(\frac{1}{r^3}-\frac{1}{r_0^3}\right)}\\
&+&\frac{r^2d\phi^2}{\exp[{\frac{2r_0}{r}}]}.
\end{eqnarray}
The Gaussian optical curvature in leading order terms is approximated as
\begin{equation}
    K \simeq \frac{ r_0}{2r^3}-\frac{r_0^2}{2 r^4}+\frac{ \pi^3(3r^3+9 r^2 r_0-14r_0^3)}{540 r^6 r_0}.
\end{equation}
Approximating this expression in leading order, the deflection angle reads
\begin{equation}
\hat{\alpha}\simeq -\int\limits_{0}^{\pi }\int\limits_{\frac{\mathsf{b}}{\sin \phi }%
}^{\infty }\left[  \frac{ r_0}{2r^3}-\frac{r_0^2}{2 r^4}+\frac{ \pi^3(3r^3+9 r^2 r_0-14r_0^3)}{540 r^6 r_0} \right]r dr d\phi. 
\end{equation}
Solving this integral we find the following solution
\begin{equation}
\hat{\alpha} \simeq -\frac{r_0}{\mathsf{b}}+\frac{\pi r_0^2}{8 \mathsf{b}^2}-\frac{\pi^3 }{90 r_0 \mathsf{b}}\left(1+\frac{3 \pi r_0}{8 \mathsf{b}}\right).\label{alpha2}
\end{equation}
One can infer from the above result that since the deflection of light is negative, it indicates that light rays in this case
always bend outward the wormhole due to the non-zero redshift function. Of course the resulting negative value should be taken as an absolute value $|\hat{\alpha} |$.

\subsection{Case with $\exp(2\Phi(r))=1+\frac{\gamma^2}{r^2}$}
In this particular case the optical metric in the equatorial plane reads
\begin{eqnarray}\notag
dt^2&=&\frac{\left(1+\frac{\gamma^2}{r^2}\right)^{-1}dr^2}{1-\frac{r_0}{r}-\frac{\pi^3}{90 r }\left(\frac{1}{r}-\frac{1}{r_0}\right)-\frac{\pi^3 D_{i} \beta}{270 r} \left(\frac{1}{r^3}-\frac{1}{r_0^3}\right)}\\
&+& \frac{r^2 d\phi^2}{1+\frac{\gamma^2}{r^2}}.
\end{eqnarray}
The Gaussian optical curvature in leading order terms is approximated as
\begin{eqnarray}\notag
    K &\simeq & -\frac{ r_0}{2r^3}+\frac{ \pi^3 (r-2r_0)}{180 r^4 r_0} \\
    &+&\gamma\left[\frac{180 r^2 r_0+(3\pi^2-270 r_0)r-4 \pi^3  r_0}{90 r^6 r_0}\right].
\end{eqnarray}
With this result in hand, in leading order the deflection angle is written as
\begin{eqnarray}
\hat{\alpha}&\simeq & -\int\limits_{0}^{\pi }\int\limits_{\frac{\mathsf{b}}{\sin \phi }%
}^{\infty }\left[ -\frac{ r_0}{2r^3}+\frac{ \pi^3 (r-2r_0)}{180 r^4 r_0} \right]r dr d\phi\\\notag
&-& \gamma \int\limits_{0}^{\pi }\int\limits_{\frac{\mathsf{b}}{\sin \phi }%
}^{\infty }\left[ \frac{180 r^2 r_0+(3\pi^2-270 r_0)r-4 \pi^3  r_0}{90 r^6 r_0} \right]r dr d\phi. 
\end{eqnarray}
Solving this integral we find the following solution
\begin{equation}
\hat{\alpha} \simeq \frac{r_0}{\mathsf{b}}-\frac{\gamma \pi}{2 \mathsf{b}^2}-\frac{\pi^3 }{90 r_0 \mathsf{b}}\left(1-\frac{\pi r_0}{4\,\mathsf{b}}\right).\label{alpha3}
\end{equation}
As was expected, in the limit $\gamma \to 0$, we recover deflection angle given by Eq. (\ref{alga}). In other words, the presence of the parameter $\gamma$ decreases the deflection angle compared to Eq. (\ref{alga}). The first and the second term are related to the geometric structure of the wormhole, while the third term encodes the semiclassical quantum effects.

\subsection{Case with $\exp(2\Phi(r))=\left(\frac{1}{1+\frac{\beta  D_i}{r^2}}\right)^{\frac{2}{1+1/\omega}}$}
\begin{figure}
\includegraphics[width=8.2cm]{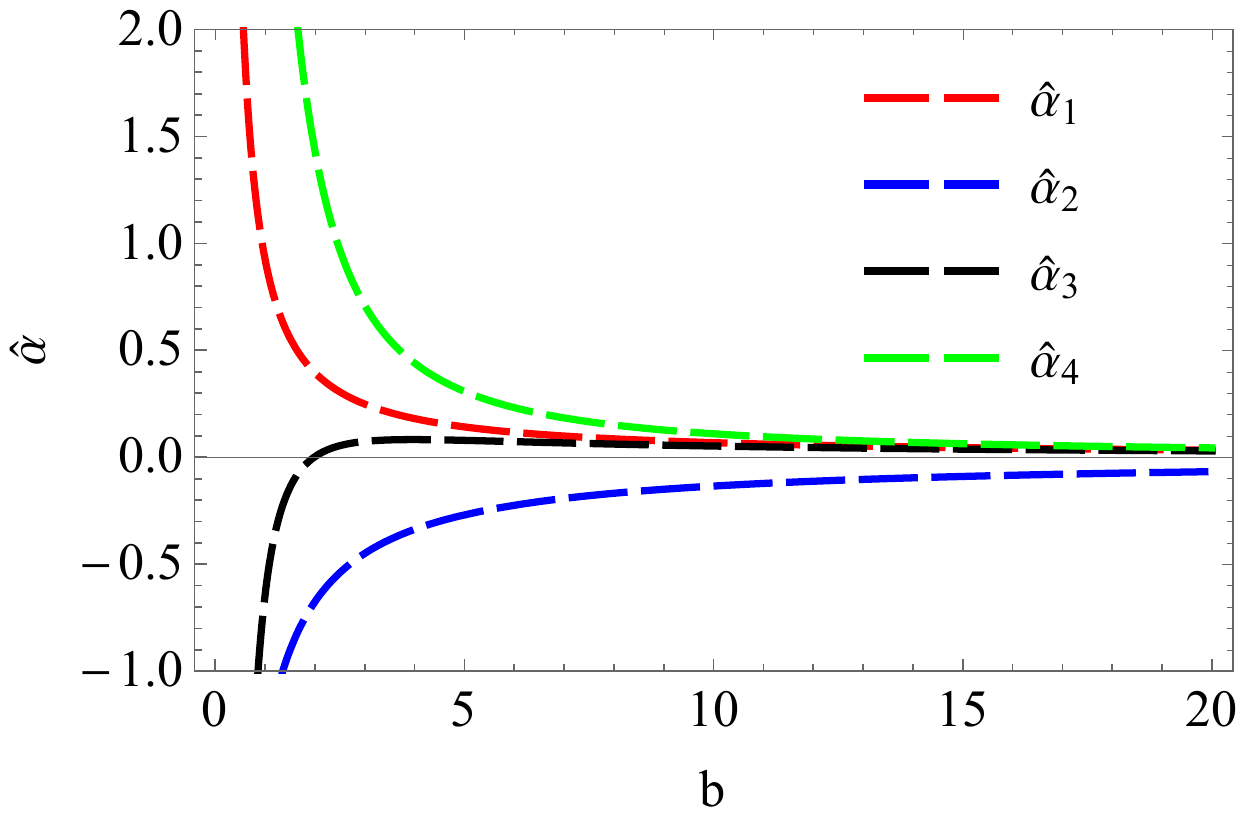}
\caption{ The deflection angle against the impact parameter $b$ using Eq. (\ref{alga}), (\ref{alpha2}), (\ref{alpha3}) and (\ref{alpha4}), respectively.  We use $r_0=1$, $\hbar=1$ and $\beta=0.1$, and $\gamma=1$ for the case $D_1$. The blue curve corresponds to Eq. (\ref{alpha2}) showing that the light rays bend outward the wormhole. On the other hand, the effect of $\gamma$ decreases the deflection angle (black curve) compared to (\ref{alga}) (red curve). The deflection angle (\ref{alpha4}), corresponds to the anisotropic wormhole (green curve). }\label{fig18}
\end{figure}

In this particular case the optical metric in the equatorial plane reads
\begin{eqnarray}\notag
dt^2&=&\frac{\left(\frac{1}{1+\frac{\beta  D_i}{r^2}}\right)^{-\frac{2}{1+1/\omega}}dr^2}{1-\frac{r_0}{r}-\frac{\pi^3}{90 r }\left(\frac{1}{r}-\frac{1}{r_0}\right)-\frac{\pi^3 D_{i} \beta}{270 r} \left(\frac{1}{r^3}-\frac{1}{r_0^3}\right)}\\
&+& \frac{r^2 d\phi^2}{\left(\frac{1}{1+\frac{\beta  D_i}{r^2}}\right)^{\frac{2}{1+1/\omega}}}.
\end{eqnarray}
Let us consider the special case with $\omega=1$. The Gaussian optical curvature in leading order terms is approximated as
\begin{eqnarray}
    K &\simeq & -\frac{ r_0}{2r^3}+\frac{ \pi^3 (r-2r_0)}{180 r^4 r_0} \\\notag
    && + \frac{\beta D_i[\pi^3 r^3-1080 r^2 r_0^3+\mathcal{J} r+20 \pi^3 r_0^3]}{540 r_0^3r^6},
\end{eqnarray}
where 
\begin{equation}
    \mathcal{J}=1620 r_0^4-18 \pi^3 r_0^2.
\end{equation}
With this result in hand, in leading order the deflection angle is written as
\begin{eqnarray}
\hat{\alpha}&\simeq & -\int\limits_{0}^{\pi }\int\limits_{\frac{\mathsf{b}}{\sin \phi }%
}^{\infty }\left[ -\frac{ r_0}{2r^3}+\frac{ \pi^3 (r-2r_0)}{180 r^4 r_0} \right]r dr d\phi\\\notag
&&-\beta D_i \int\limits_{0}^{\pi }\int\limits_{\frac{\mathsf{b}}{\sin \phi }%
}^{\infty }\frac{[\pi^3 r^3-1080 r^2 r_0^3+\mathcal{J} r+20 \pi^3 r_0^3]}{540 r_0^3r^6}r dr d\phi.
\end{eqnarray}
Solving this integral we find the following solution
\begin{equation}
\hat{\alpha} \simeq \frac{r_0}{\mathsf{b}}+\frac{\beta D_i \pi }{2 \mathsf{b}^2}-\frac{\pi^3 }{90 r_0 \mathsf{b}}\left(1-\frac{\pi r_0}{4\,\mathsf{b}}\right).\label{alpha4}
\end{equation}
In this case, beside the first term which is related to the wormhole geometry, we find an effect of GUP parameter $\beta$ in leading order terms on the deflection angle encoded in the second term, while the third term is related to the semiclassical quantum effects. We show graphically the dependence of deflection angle against the impact parameter in Fig. (\ref{fig18}).
\section{Conclusion}
In this paper, we have explored the effect of the Generalized Uncertainty Principle (GUP)  on the Casimir wormhole spacetime. In particular, we have constructed three types of the GUP relations, namely the KMM model, DGS model, and finally the so called type II model for GUP principle. To this end, we have used three different models of the redshift function, i.e., $\Phi(r)=constant$, along with $\Phi(r)=r_0/r$ and $\exp(2\Phi(r))=1+\frac{\gamma^2}{r^2}$, to obtain a class of asymptotically flat wormhole solutions supported by Casimir energy under the effect of GUP.  Having used the specific model for the wormhole geometry, we then used two EoS models $\mathcal{P}_r(r)=\omega_{r}(r) \rho(r)$ and $\mathcal{P}_t(r)=\omega_{t}(r)\mathcal{P}_r(r)$ to obtained the specific relation for the EoS parameter $\omega_{r}(r)$ and $\omega_{t}(r)$, respectively. In addition, we have considered the isotropic wormhole and found an interesting solution describing an asymptotically flat GUP wormhole with anisotropic matter. 

Furthermore, we have checked the null, weak, and strong conditions at the wormhole throat with a radius $r_0$, and shown that in general the classical energy conditions are violated by some small and arbitrary quantities at the wormhole throat. However, we have also highlighted the Quantum Weak Energy Condition (QWEC) according to which such small violations are possible due to the quantum fluctuations. In this direction, we have also examined the ADM mass of the wormhole and the volume integral quantifier to calculate the amount of the exotic matter near the wormhole throat, such that the wormhole extends form $r_0$ to a cut off radius located at $`a'$. We studied the embedding diagram to show that with the increase of the GUP parameter there is an effect on the effective geometry of the GUP wormhole.

Finally, we have used the GBT to obtain the deflection angle in three wormhole geometries. We argued that the deflection angle in leading order terms is affected by the semiclassical quantum effect as well as the wormhole throat radius. As an interesting observation, we have found that the choice of the redshift function plays a significant role in determining the deflection angle. For example, in the case $\Phi=constant$ and  $\exp(2\Phi(r))=1+\frac{\gamma^2}{r^2}$ the light rays bend towards the wormhole, while in contrast having $\Phi(r)=r_0/r$, we discovered that light rays bent outward the wormhole. We also found that the deflection angle depends upon the parameter $\gamma$, while there is an effect of the GUP parameter in leading order terms only in the case of anisotropic GUP wormhole. However, a thorough analysis of these effects will be intentionally left for further investigation.

\end{document}